\documentclass[a4paper,11pt]{article}
\usepackage{stmaryrd}
\usepackage{amsfonts}
\usepackage{bbm}
\usepackage{amscd}
\usepackage{mathrsfs}
\usepackage{latexsym,amssymb,amsmath,amscd,amscd,amsthm,amsxtra,xypic}
\usepackage[dvips]{graphicx}
\usepackage[utf8]{inputenc}
\usepackage[T1]{fontenc}
\usepackage{lmodern}
\usepackage{amssymb}
\usepackage[all]{xy}
\usepackage{nicefrac,mathtools,enumitem}
\usepackage{microtype}

\newtheorem{thm}{Theorem}[section]
\newtheorem{lem}[thm]{Lemma}
\newtheorem{cor}[thm]{Corollary}
\newtheorem{pro}[thm]{Proposition}

\newtheorem{rmk}[thm]{Remark}
\newtheorem{defi}[thm]{Definition}

\newcommand{\be }{\begin{equation}}
\newcommand{\ee }{\end{equation}}

\newcommand{\pf}{\noindent{\bf Proof.}\ }

\newcommand{\huaA}{\mathcal{A}}

\newcommand{\huaE}{\mathcal{E}}

\newcommand{\huaG}{\mathcal{G}}

\newcommand{\huaC}{{\mathcal{C}}}
\newcommand{\huaD}{\mathcal{D}}

\newcommand{\huaJ}{\mathcal{J}}

\newcommand{\frkf}{\mathfrak f}
\newcommand{\frkg}{\mathfrak g}

\newcommand{\frkp}{\mathfrak p}

\newcommand {\emptycomment}[1]{{}}

\newcommand{\frkX}{\mathfrak X}

\def\qed{\hfill ~\vrule height6pt width6pt depth0pt}

\newcommand{\half}{\frac{1}{2}}

\newcommand{\ppair}[1]{\left ( #1\right )_+}

\newcommand{\Courant}[1]{\left\llbracket  #1\right\rrbracket }
\newcommand{\Dorfman}[1]{\{ #1\}}

\newcommand{\br}[1]{   [ \cdot,    \cdot  ]   }
\newcommand{\id}{\rm{id}}

\newcommand{\dM}{\mathrm{d}}

\newcommand{\Hom}{\mathrm{Hom}}

\newcommand{\Ker}{\mathrm{ker}}

\newcommand{\ad}{\mathrm{ad}}

\newcommand{\Img}{\mathrm{Im}}

\begin{document}
\title{
{The Pontryagin Class for  Pre-Courant Algebroids
\thanks
 {
Research supported by NSFC (11101179,11471139) and NSF of Jilin Province (20140520054JH). Xiaomeng Xu is supported by the grant ERC project MODFLAT and the NCCR SwissMap of the Swiss National Science Foundation.
 }
} }
\author{\vspace{3mm}  Zhangju Liu$^a$,~~Yunhe Sheng$^b$ ~~ and ~~ Xiaomeng Xu$^c$\\
$^a $Department of Mathematics and LMAM, Peking University,
\\\vspace{3mm} Beijing
100871, China\\
$^b $Department of Mathematics, Jilin University,\\\vspace{3mm}
 Changchun 130012,  China
\\
$^c$Section of Mathematics, University of Geneva \\\vspace{3mm}
2-4 Rue de Li${\rm\grave{e}}$vre, c.p. 64, 1211-Gen${\rm\grave{e}}$ve
4, Switzerland\\
email: liuzj@pku.edu.cn; ~shengyh@jlu.edu.cn;~
xiaomeng.xu@unige.ch}
\date{}
 \maketitle

\begin{abstract}
In this paper, we show that the Jacobiator $J$ of a pre-Courant
algebroid is closed naturally. The corresponding  equivalence class
$[J^\flat]$ is defined as the Pontryagin class, which is the
obstruction  of  a pre-Courant algebroid to be deformed into a
Courant algebroid. We construct a Leibniz 2-algebra  and a Lie
2-algebra associated to a pre-Courant algebroid and prove that these
algebraic structures are isomorphic under deformations. Finally, we
introduce the twisted action of a Lie algebra on a manifold to give
more examples of pre-Courant algebroids, which include the Cartan
geometry.

\end{abstract}
\tableofcontents
\section{Introduction}
Recently, people have paid more attention to higher categorical
structures by reasons in both mathematics and physics. One way to
provide higher categorical structures is by categorifying existing
mathematical concepts. One of the simplest higher structures is a
$2$-vector space, which is the  categorification of a vector space.
If we further put a compatible Lie algebra structure on a $2$-vector
space, then we obtain a Lie $2$-algebra \cite{baez:2algebras}. The
Jacobi identity is replaced by a natural transformation, called the
Jacobiator, which also satisfies some coherence laws of its own.
Recently, the relation among higher categorical structures and
multisymplectic structures, Courant algebroids, and Dirac structures
are  studied in \cite{baezrogers,Leibniz2,zambon}.

A $2$-vector space is equivalent to a $2$-term complex of vector
spaces. A Lie $2$-algebra is equivalent to a $2$-term
$L_\infty$-algebra. $L_\infty$-algebras,  sometimes  called strongly
homotopy (sh) Lie algebras,  were introduced in \cite{stasheff2} as
a model for ``Lie algebras that satisfy Jacobi identity up to all
higher homotopies''. The notion of Leibniz algebras was introduced
by Loday in \cite{Loday and Pirashvili}, which is a generalization of Lie algebras. Their crossed
modules were also introduced in the same paper to study
the cohomology of Leibniz algebras. As a  model for ``Leibniz
algebras that satisfy Jacobi identity up to all higher homotopies'',
  the notion of strongly homotopy (sh)
Leibniz algebra, or $Lod_\infty$-algebra was given in \cite{livernet} by Livernet,
 which is further studied by Ammar, Poncin and Uchino
in \cite{ammardefiLeibnizalgebra,UchinoshL}. In \cite{Leibniz2}, the author introduced the
notion of Leibniz 2-algebras, and proved that the category of
Leibniz 2-algebras is equivalent to the category of 2-term sh
Leibniz algebras.

Courant algebroid was introduced in \cite{lwx} to study the double
of Lie bialgebroids. Equivalent definition was given in
\cite{Roytenbergthesis}. Courant algebroids have been widely studied
because of their applications in both mathematics and physics
\cite{CA,GualtieriGeneralizedComplex,Yvettehistory}. Roytenberg and Weinstein proved that every
Courant algebroid gives rise to a  Lie 2-algebra
\cite{roytenbergshl}.
Recently, two kinds of generalizations
of Courant algebroids are studied. One generalization is letting the
pseudo metric taking values in a section space of a vector bundle
instead of $C^\infty(M)$, see \cite{CLSecourant,GS,Jot,Libland} for more
details along this direction. The other generalization is relaxing
the restriction of Jacobi identity, which includes twisted Courant
algebroids by closed 4-forms, $H$-twisted Courant algebroids as well
as pre-Courant algebroids. Just to list a few of the work in this direction, Hansen and Strobl introduced the notion
of twisted Courant algebroids by closed $4$-forms in \cite{4form},
which arises from the study of three dimensional sigma models with
Wess-Zumino term.  In \cite{Leibniz2}, the authors proved that
associated to any twisted Courant algebroid by a closed $4$-form, there is a Leibniz
2-algebra. In general, if one studies generalized geometry, this
$4$-form will arise naturally as background \cite{Hull}. The construction of twisted Courant algebroids from coisotropic Cartan geometry is given in \cite{xiaomeng}. In
\cite{Grutzmann, GrutzmannC}, Gr{\"u}tzmann introduced $H$-twisted
Lie (Courant) algebroids, in which the Jacobiator is controlled by
an $H\in\Hom(\wedge^3E,\Ker(\rho))$, which satisfies some ``closed''
condition itself.  The notions of pre-Courant algebroids and Courant vector bundles were
introduced by Vaisman in \cite{Vai05}, and its relation to parabolic geometry was
studied in \cite{parabolic}.

In this paper, we find that the Jacobiator $J\in \huaC^3(E,\ker(\rho))$ in a pre-Courant
algebroid $(E,(\cdot,\cdot)_+,\rho,\circ)$ is totally skew-symmetric, and satisfies some ``closed''
condition naturally, i.e. $\partial J=0$. Equivalently, $\huaD J^\flat=0$, where $J^\flat\in C^4(E)$ is defined by $J^\flat(e_1,e_2,e_3,e_4)=(J(e_1,e_2,e_3),e_4)_+$. Thus, it is not necessary to assume that the
Jacobiator is controlled by something in advance. In fact, we showed
in this paper that  {\bf $\mbox{pre-CA}=H\mbox{-twisted~ CA}.$}
Since  twisted Courant algebroids by closed $4$-forms arise from the
study of three dimensional sigma models with Wess-Zumino term
naturally, it will be interesting to study the physical meaning of a
general pre-Courant algebroid.  The Pontryagin class for a quadratic Lie algebroid $\huaA$ was introduced in \cite{dissection} as the obstruction of $\huaA$ to be the ample Lie algebroid of a Courant algebroid. Please also see \cite{firstP} for more information about the Pontryagin class. We define the equivalence class $[J^\flat]$ as the Pontryagin class for a pre-Courant algebroid, and show that it is the obstruction to be deformed into a Courant algebroid.

 It
turns out that, one can obtain a Leibniz 2-algebra as well as a Lie
2-algebra associated to a pre-Courant algebroids. It is surprised
that the Leibniz 2-algebra associated to the deformed pre-Courant
algebroid is isomorphic to the Leibniz 2-algebra associated to the
original one. Since the closed 4-form is exact in an exact twisted
Courant algebroid, it turns out that the Leibniz 2-algebra
associated to the exact twisted Courant algebroid by a closed 4-form
is isomorphic to the strict Leibniz algebra associated to the standard Courant
algebroid.

It is known that  a coisotropic action of a quadratic Lie algebra on
a manifold could give a Courant algebroid. We introduce the notion
of a twisted action. A coisotropic twisted action of a quadratic Lie
algebra on a manifold gives rise to a pre-Courant algebroid. We also
study transitive pre-Courant algebroids in detail.

The paper is organized as follows. In Section 2, we recall some
basic notions of Leibniz cohomologies, Leibniz 2-algebras, morphisms
between Leibniz 2-algebras and Lie 2-algebras. In Section 3, for a
pre-Courant algebroid $(E,\ppair{\cdot,\cdot},\rho,\circ)$, we
introduce the covariant derivatives $\partial$ and $\huaD$ using the
coboundary operator for Leibniz cohomology and Lie algebroid
cohomology respectively.  We prove that the Jacobiator $J\in
\huaC^3(E,\ker(\rho))$, and satisfies $\partial J=0$ (Theorem
\ref{thm:J0}). In Section 4, by the fact that $\partial J=0$, we
obtain a Leibniz 2-algebra associated to a pre-Courant algebroid
(Theorem \ref{thm:Leibniz2}). Moreover, by using the skew-symmetric
bracket, we can obtain a Lie 2-algebra (Theorem \ref{thm:Lie3}).
Finally, we prove that  the Leibniz 2-algebras associated to all
pre-Courant algebroid structures on a Courant vector bundle
$(E,\ppair{\cdot,\cdot},\rho)$ are isomorphic. In Section 5, we show
that a transitive pre-Courant algebroid satisfying $\Img
J\subset\rho^*(T^*M)$ is exactly a twisted Courant algebroid by a
closed 4-form. Since in this special case, the closed 4-form is
known as the Pontryagin class of the induced ample Lie algebroid
\cite{firstP,dissection, Leibniz2}, we call the equivalence class of
$J^\flat$ the Pontryagin class of the pre-Courant algebroid.  In
Section 6, we introduce the notion of twisted action, by which we
construct pre-Courant algebroids. Examples arising from Cartan
geometry \cite{weylstructure} are given.

\section{Preliminaries}
In this section, we recall Leibniz algebras, Leibniz 2-algebras and
 Lie 2-algebras.

A Leibniz algebra $\frkg$ is an $R$-module, where $R$ is a
commutative ring, endowed with a linear map
$[\cdot,\cdot]_\frkg:\frkg\otimes\frkg\longrightarrow\frkg$
satisfying
$$
[g_1,[g_2,g_3]_\frkg]_\frkg=[[g_1,g_2]_\frkg,g_3]_\frkg+[g_2,[g_1,g_3]_\frkg]_\frkg,\quad
\forall~g_1,g_2,g_3\in \frkg.
$$

This is in fact a left Leibniz algebra. In this paper, we only
consider left Leibniz algebras.
 Recall that a representation of the Leibniz algebra
$(\frkg,[\cdot,\cdot]_\frkg)$ is an $R$-module $V$ equipped with,
respectively, left and right actions  of $\frkg$ on $V$,
$$[\cdot,\cdot]:\frkg\otimes V\longrightarrow V,\quad [\cdot,\cdot]:V\otimes\frkg \longrightarrow V,$$
such that for any $g_1,g_2\in\frkg$, the following equalities hold:
\begin{equation}\label{condition of rep}
l_{[g_1,g_2]}=[l_{g_1},l_{g_2}],\quad
r_{[g_1,g_2]}=[l_{g_1},r_{g_2}],\quad r_{g_2}\circ
l_{g_1}=-r_{g_2}\circ r_{g_1},
\end{equation} where
$l_{g_1}u=[g_1,u]$ and $r_{g_1}u=[u,g_1]$ for any $u\in V$. The
Leibniz cohomology of $\frkg$ with coefficients in $V$ is the
homology of the cochain complex
$C^k(\frkg,V)=\Hom_R(\otimes^k\frkg,V), (k\geq0)$ with the
coboundary operator $\partial:C^k(\frkg,V)\longrightarrow
C^{k+1}(\frkg,V)$ defined by
\begin{eqnarray}
\nonumber\partial c^k(g_1
,\dots,g_{k+1})&=&\sum_{i=1}^k(-1)^{i+1}l_{g_i}(c^k(g_1,\dots,\widehat{g_i},\dots,g_{k+1}))
+(-1)^{k+1}r_{g_{k+1}}(c^k(g_1,\dots,g_k))\\
\label{formulapartial}&&+\sum_{1\leq i<j\leq
k+1}(-1)^ic^k(g_1,\dots,\widehat{g_i},\dots,g_{j-1},[g_i,g_j]_\frkg,g_{j+1},\dots,g_{k+1}).
\end{eqnarray}
The fact that $\partial\circ\partial=0$ is proved in \cite{Loday and
Pirashvili}. \vspace{3mm}

The notion of strongly homotopy (sh) Leibniz algebras, or
$Lod_\infty$-algebras was introduced in \cite{livernet}. See also \cite{ammardefiLeibnizalgebra,UchinoshL} for more
details. Here we only consider the 2-term case. Leibniz 2-algebras
were introduced in \cite{Leibniz2} as the categorification of
Leibniz algebras. The category of 2-term sh Leibniz algebras is
equivalent to the category of Leibniz 2-algebras. Due to this
reason, a 2-term sh Leibniz algebra will be called a Leibniz
2-algebra in the sequel.

\begin{defi}\label{defi:2leibniz}
  A Leibniz $2$-algebra $\mathbb V$ consists of the following data:
\begin{itemize}
\item[$\bullet$] a complex of vector spaces $\mathbb V:V_1\stackrel{\dM}{\longrightarrow}V_0,$

\item[$\bullet$] a bilinear map $l_2:V_i\times V_j\longrightarrow
V_{i+j}$, where $i+j\leq 1$,

\item[$\bullet$] a  trilinear map $l_3:V_0\times V_0\times V_0\longrightarrow
V_1$,
   \end{itemize}
   such that for any $w,x,y,z\in V_0$ and $m,n\in V_1$, the following equalities are satisfied:
\begin{itemize}
\item[$\rm(a1)$] $\dM l_2(x,m)=l_2(x,\dM m),$
\item[$\rm(a2)$]$\dM l_2(m,x)=l_2(\dM m,x),$
\item[$\rm(a3)$]$l_2(\dM m,n)=l_2(m,\dM n),$
\item[$\rm(b1)$]$\dM l_3(x,y,z)=l_2(x,l_2(y,z))-l_2(l_2(x,y),z)-l_2(y,l_2(x,z)),$
\item[$\rm(b2)$]$ l_3(x,y,\dM m)=l_2(x,l_2(y,m))-l_2(l_2(x,y),m)-l_2(y,l_2(x,m)),$
\item[$\rm(b3)$]$ l_3(x,\dM m,y)=l_2(x,l_2(m,y))-l_2(l_2(x,m),y)-l_2(m,l_2(x,y)),$
\item[$\rm(b4)$]$ l_3(\dM m,x,y)=l_2(m,l_2(x,y))-l_2(l_2(m,x),y)-l_2(x,l_2(m,y)),$
\item[$\rm(c)$] the Jacobiator identity:\begin{eqnarray*}
&&l_2(w,l_3(x,y,z))-l_2(x,l_3(w,y,z))+l_2(y,l_3(w,x,z))+l_2(l_3(w,x,y),z)\\
&&-l_3(l_2(w,x),y,z)-l_3(x,l_2(w,y),z)-l_3(x,y,l_2(w,z))\\
&&+l_3(w,l_2(x,y),z)+l_3(w,y,l_2(x,z))-l_3(w,x,l_2(y,z))=0.\end{eqnarray*}
   \end{itemize}
\end{defi}
We usually denote a 2-term sh Leibniz algebra by
$(V_1\stackrel{\dM}{\longrightarrow}V_0,l_2,l_3)$, or simply by
$\mathbb V$. In particular, if $l_2$ and $l_3$ are skew-symmetric,
we obtain a 2-term $L_\infty$-algebra, which is also called a {\bf
Lie 2-algebra}.

\begin{defi}\label{defi:Leibniz morphism}
 Let $\mathbb V$ and $\mathbb V^\prime$ be  Leibniz $2$-algebras, a morphism $\frkf$
 from $\mathbb V$ to $\mathbb V^\prime$ consists of
\begin{itemize}
  \item[$\bullet$] linear maps $f_0:V_0\longrightarrow V_0^\prime$
  and $f_1:V_1\longrightarrow V_1^\prime$ commuting with the
  differential, i.e.
  $$
f_0\circ \dM=\dM^\prime\circ f_1;
  $$
  \item[$\bullet$] a bilinear map $f_2:V_0\times V_0\longrightarrow
  V_1^\prime$,
\end{itemize}
  such that  for all $x,y,z\in L_0,~m\in
L_1$, we have
\begin{equation}\label{eqn:DGLA morphism c 1}\left\{\begin{array}{rll}
l_2^\prime(f_0(x),f_0(y))-f_0l_2(x,y)&=&\dM^\prime
f_2(x,y),\\
l_2^\prime(f_0(x),f_1(m))-f_1l_2(x,m)&=&f_2(x,\dM
m),\\
l_2^\prime(f_1(m),f_0(x))-f_1l_2(m,x)&=&f_2(\dM
m,x),\end{array}\right.
\end{equation}
and
\begin{eqnarray}
\nonumber&&f_1(l_3(x,y,z))+l_2^\prime(f_0(x),f_2(y,z))-l_2^\prime(f_0(y),f_2(x,z))-l_2^\prime(f_2(x,y),f_0(z))\\
\label{eqn:DGLA morphism c 2}
&&-f_2(l_2(x,y),z)+f_2(x,l_2(y,z))-f_2(y,l_2(x,z))-l_3^\prime(f_0(x),f_0(y),f_0(z))=0.
\end{eqnarray}
\end{defi}

If $(f_0,f_1)$ is an isomorphism of underlying complexes, we say
that $(f_0,f_1,f_2)$ is an isomorphism.

\section{Pre-Courant algebroids}

Let $(E,\ppair{\cdot,\cdot})$ be a pseudo-Euclidean vector bundle over $M$. The pseudo metric $\ppair{\cdot,\cdot}$ induces an isomorphism $\Xi$ from $E$ to $E^*$  via $\Xi(e_1)(e_2)=\ppair{e_1,e_2}$. A {\it Courant vector bundle} is a pseudo-Euclidean vector bundle
$(E,\ppair{\cdot,\cdot})$ with an anchor  $\rho:
E\longrightarrow{TM}$, such that $\rho \rho^*= 0$, where $ \rho^*:
T^*M\longrightarrow{E^*} \cong E$ is the dual map of $\rho$. See \cite{Vai05} for more details. With the identification  above,
it is easy to see that $(\ker\rho)^\perp$ coincides with
$\rho^*(T^*M)$. That is, $\ker(\rho)$ is a co-isotropic distribution
in $E$.
\begin{defi}{\rm\cite{Vai05}}\label{defi:preCA}
A pre-Courant algebroid is a Courant  vector bundle
$(E,\ppair{\cdot,\cdot},\rho)$ with an operation ``$\circ$'' on
$\Gamma(E)$ satisfying:
\begin{itemize}
\item[ \rm   (i)] $\rho({e_1}\circ{e_2})=[\rho(e_1),\rho(e_2)]$;
\item[\rm   (ii)] $\ppair{e_{1}\circ{e_{1}}, e_2}    = \half\rho(e_2)\ppair{e_{1},e_{1}}$;
\item[\rm    (iii)]
$\rho(e_{1})\ppair{e_{2},e_{3}}=\ppair{e_{1}\circ{e_{2}},e_{3}}+\ppair{e_{2},e_{1}\circ{e_{3}}}$,
\, \, $\forall e_1,e_2,e_3\in{\Gamma(E)}$.
\end{itemize}
\end{defi}
Define $J:\Gamma(E)\times \Gamma(E)\times \Gamma(E)\longrightarrow \Gamma(E)$ by
$$
J(e_1,e_2,e_3)=
e_{1}\circ(e_{2}\circ{e_{3})}-(e_{1}\circ{e_{2}})\circ{e_{3}}-e_{2}\circ(e_{1}\circ{e_{3}}),
$$
which is called the {\bf Jacobiator} of the pre-Courant algebroid structure. So far
some special cases for the Jacobiator were studied more or less. $J
= 0$ is just a Courant algebroid \cite{lwx,Roytenbergthesis}. If
there is a closed 4-form $H\in\Omega^4(M)$ such that
$$ J(e_1,e_2,e_3)
=\rho^*(i_{\rho(e_1)\wedge\rho(e_2)\wedge\rho(e_3)}H),
$$
 one gets a twisted Courant algebroid by a closed 4-form $H$, which arise naturally
from the study of three dimensional sigma models with Wess-Zumino
term (\cite{4form}, \cite{Leibniz2}).

For a Courant vector bundle, one can define a differential operator:
\begin{equation}\label{eqn:defi D}
\huaD:C^\infty(M)\rightarrow\Gamma(E)\, \, by \, \,
\ppair{\huaD f,e}:=\rho(e)f.\end{equation} Then   (ii) in Definition \ref{defi:preCA} is
equivalent to
\begin{equation}\label{eqn:skew}
  e_1\circ e_2+e_2\circ e_1=\huaD\ppair{e_1,e_2}.
\end{equation}
and one can easily get the following properties for a pre-Courant
algebroid, which are the same as the ones for a Courant algebroid.
\begin{lem}\label{lem:rep1}
Let  $(E,\ppair{\cdot,\cdot},\rho,\circ)$ be a pre-Courant
algebroid. For all ${e_1,e_2,e}\in{\Gamma(E)}$ and
$f\in{C^\infty(M)}$, we have:
\begin{eqnarray}
\label{pro1}e_1\circ(fe_2)&=&f(e_1\circ{e_2})+(\rho(e_1)f)e_2, \\
\label{pro2}(fe_1)\circ(e_2)&=&f(e_1\circ{e_2})-(\rho(e_2)f){e_1}+\ppair{e_1,e_2}\huaD{f},\\
\label{pro3}(\huaD{f})\circ{e}&=&0,\\
\label{pro3'}
e\circ(\huaD{f})&=&{\huaD}(\rho(e)f),\\
\label{pro4}\rho\circ\huaD&=&0.
\end{eqnarray}
\end{lem}

To explain  that the Jacobiator $J$ in a pre-Courant algebroid is closed in some way, we introduce the following   ``quasi-cochain complex''.
Similar to \cite{Lodayalg}, for a pre-Courant algebroid $E$, let $C^k (E)$ be a subspace of the section space $\Gamma(\wedge^k E^*)$ as follows:
\begin{equation}
C^k (E)=\{\psi\in\Gamma(\wedge^k E^*)|\, i_{\huaD
f}\psi=0,~\forall f\in C^\infty(M)\}.
\end{equation}

Define  the covariant derivative $\huaD: \Gamma(\wedge^{k}E^*)\longrightarrow\Gamma(\wedge^{k+1}E^*)
$ by
\begin{eqnarray}
 \nonumber \huaD\psi(e_1,\cdots,e_{k+1})&=&\sum_{i=1}^{k+1}(-1)^{i+1}\rho(e_i)
  \psi(e_1,\cdots,\hat{e_i},\cdots,e_{k+1})\\
\label{eqn:defihuaD} &&+\sum_{i<j}(-1)^{i+j}\psi(e_i\circ
 e_j,e_1,\cdots,\hat{e_i},\cdots,\hat{e_j},\cdots,e_{k+1}),
\end{eqnarray}
which  extends  the differential operator given in \eqref{eqn:defi D}.
 By \eqref{pro3} and \eqref{pro3'}, it is easy to see that
$\huaD( C^k (E))\subset C^{k+1} (E)$. But, in general, $(C^{\bullet} (E), \, \huaD)$ is not a cochain complex since $\huaD^2 \neq 0$  by the
 nontrivial Jacobiator. An equivalent  quasi-cochain complex  $(\huaC^{\bullet} (E, \ker(\rho)), \, \partial)$ is given by: a bundle map
$\phi : \wedge^{k}E \rightarrow E$  is in $\huaC^k (E,\ker(\rho))$ if and only if

\begin{itemize}
  \item[\rm (1)] $\Img\phi  \subset\ker(\rho)$;
  \item[\rm(2)] $i_{\huaD f}\phi=0, $~~\, for all $f\in{C^\infty(M)}$;
 \item[\rm(3)] The map $(e_1,\cdots,e_k,e_{k+1})\longmapsto (\phi(e_1,\cdots,e_k),e_{k+1})_+$
is totally skew-symmetric.
\end{itemize}
It is obvious that an element  $\phi\in \huaC^k(E,\ker(\rho))$ induces an element $\phi^\flat\in
C^{k+1}(E)$ via
$$
\phi^\flat(e_1,\cdots,e_{k+1})=\ppair{\phi(e_1.\cdots,e_{k}),e_{k+1}}.
$$
Similarly, one can define a covariant derivative
$$\partial:\huaC^k (E,\ker(\rho))\longrightarrow \huaC^{k+1} (E,\ker(\rho)) \subset \Gamma(\Hom(\wedge^{k+1}E,E))$$ by
\begin{eqnarray}
 \nonumber \partial\phi(e_1,\cdots,e_{k+1})&=&\sum_{i=1}^k(-1)^{i+1}e_i\circ
  \phi(e_1,\cdots,\hat{e_i},\cdots,e_{k+1})\\
  \nonumber&&+(-1)^{k+1}\phi(e_1,\cdots,e_{k})\circ e_{k+1}\\
 \nonumber &&+\sum_{i<j}(-1)^{i+j}\phi(e_i\circ e_j,e_1,\cdots,\hat{e_i},\cdots,\hat{e_j},\cdots,e_{k+1}).
\end{eqnarray}
It is straightforward to see that
$\partial \phi\in\huaC^{k+1} (E,\ker(\rho))$.
About the relation between two covariant derivatives $\huaD$ and $\partial$ given above, we have

\begin{lem}\label{lem:comm}
 With the notations above, for all $\phi \in \huaC^{k}(E,\ker(\rho))$, we have $$\huaD(\phi^\flat) = (\partial
 \phi)^\flat.$$
 \end{lem}
\pf By (iii) in Definition \ref{defi:preCA} and \eqref{eqn:skew}, after a straightforward computation, we get
\begin{eqnarray*}
   (\partial \phi)^\flat(e_1,\cdots,e_{k+1})
  &=&\huaD( \phi^\flat)(e_1,\cdots,e_{k+1}),
\end{eqnarray*}
which implies that $\huaD(\phi^\flat) = (\partial
 \phi)^\flat.$
\qed\vspace{3mm}

The follows is the main result of this section.
\begin{thm}\label{thm:J0}
Let  $(E,\ppair{\cdot,\cdot},\rho,\circ)$ be a pre-Courant
algebroid. Then the Jacobiator $J\in \huaC^{ 3}(E,\ker(\rho))$ and
 $\partial J=0$. Or equivalently, $J^\flat \in C^4 (E)$ and
$\huaD (J^\flat)=0$.
\end{thm}

\begin{lem}\label{lem:Jskew}
  The Jacobiator $J$ is skew-symmetric.
\end{lem}
\pf  By \eqref{pro3}, we have
\begin{eqnarray*}
J(e_1,e_2,e_3)+J(e_2,e_1,e_3)=-2\huaD{\ppair{e_1,e_2}}\circ{e_3}=0.
\end{eqnarray*}
By \eqref{pro3'} and   (iii) in Definition \ref{defi:preCA}
\begin{eqnarray*}
J(e_1,e_2,e_3)+J(e_1,e_3,e_2)&=&2e_1\circ\huaD{\ppair{e_2,e_3}}-2\huaD{\ppair{e_2,e_1\circ{e_3}}}-2\huaD{\ppair{e_1\circ{e_2},e_3}}\\
&=&2e_1\circ\huaD{\ppair{e_2,e_3}}-2\huaD\big(\rho(e_1)\ppair{e_2,e_3}\big)\\
&=&0.
\end{eqnarray*}
Therefore,  the Jacobiator $J$ is skew-symmetric.\qed

\begin{lem}\label{lem:Jlinear}
The Jacobiator $J$ is $C^\infty$-linear.
\end{lem}
\pf Since $J$ is skew-symmetric, we only need to show
$$
J(e_1,e_2,fe_3)=fJ(e_1,e_2,e_3),\quad \forall f\in C^\infty(M).
$$
By   (i) in Definition \ref{defi:preCA}, we have
\begin{eqnarray*}
 J(e_1,e_2,fe_3)&=&e_{1}\circ(e_{2}\circ{fe_{3})}-(e_{1}\circ{e_{2}})\circ{fe_{3}}-e_{2}\circ(e_{1}\circ{fe_{3}})\\
 &=&e_1\circ (fe_2\circ
 e_3+\rho(e_2)(f)e_3)-f(e_{1}\circ{e_{2}})\circ{e_{3}}-\rho(e_1\circ
 e_2)(f)e_3\\
 &&-e_2\circ(fe_1\circ e_3+\rho(e_1)(f)e_3)\\
 &=&fe_{1}\circ(e_{2}\circ{e_{3})}+\rho(e_1)(f)e_2\circ
 e_3+\rho(e_2)(f)e_1\circ e_3+\rho(e_1)\rho(e_2)(f)e_3\\
 &&-f(e_{1}\circ{e_{2}})\circ{e_{3}}-[\rho(e_1),
 \rho(e_2)](f)e_3\\
 &&-fe_{2}\circ(e_{1}\circ{fe_{3}})-\rho(e_2)(f)e_1\circ
 e_3-\rho(e_1)(f)e_2\circ e_3-\rho(e_2)\rho(e_1)(f)e_3\\
 &=&fJ(e_1,e_2,e_3).
\end{eqnarray*}
Therefore, $J$ is $C^\infty$-linear. \qed

\begin{lem}\label{lem:JDf}
  For all $f\in C^\infty(M)$, $J(\huaD f,\cdot,\cdot)=0$.
\end{lem}
\pf It can be obtained directly by \eqref{pro3}. \qed\vspace{3mm}

{\bf Proof of Theorem \ref{thm:J0}} By Lemma \ref{lem:Jskew},
\ref{lem:Jlinear}, \ref{lem:JDf}, to see $J\in
\huaC^3(E,\ker(\rho))$, we only need to show that the map
$(e_1,e_2,e_3,e_4)\longmapsto\ppair{J(e_1,e_2,e_3),e_4}$ is totally
skew-symmetric. Since $J$ is skew-symmetric, we only need to show
\begin{eqnarray}
\ppair{J(e_1,e_2,e_3),e_4}+\ppair{J(e_4,e_2,e_3),e_1}&=&0.
\end{eqnarray}
The left hand side, which we denote by $L$, is equal to
\begin{eqnarray}
\label{10}\nonumber{L}&=&\ppair{e_1\circ(e_2\circ{e_3}),e_4}-\ppair{(e_1\circ{e_2})\circ{e_3},e_4}-\ppair{e_2\circ(e_1\circ{e_3}),e_4}\\
\nonumber&&+\ppair{e_4\circ(e_2\circ{e_3}),e_1}-\ppair{(e_4\circ{e_2})\circ{e_3},e_1}-\ppair{e_2\circ(e_4\circ{e_3}),e_1}.
\end{eqnarray}
Note that
\begin{eqnarray*}
  &&-\ppair{(e_1\circ{e_2})\circ{e_3},e_4}-\ppair{(e_4\circ{e_2})\circ{e_3},e_1}\\
&=&\ppair{e_3\circ(e_1\circ e_2),e_4}-\rho(e_4)\ppair{e_3,e_1\circ
e_2}+\ppair{e_3\circ(e_4\circ e_2),e_1}-\rho(e_1)\ppair{e_3,e_4\circ
e_2}.
\end{eqnarray*}
Therefore, we have
\begin{eqnarray*}
\nonumber L&=&\rho(e_1)\ppair{e_2\circ{e_3},e_4}-\rho(e_4)\ppair{e_1\circ{e_2},e_3}+\rho(e_3)\ppair{e_1\circ{e_2},e_4}\\
\nonumber&&-\rho(e_2)\ppair{e_1\circ{e_3},e_4}+\rho(e_4)\ppair{e_2\circ{e_3},e_1}-\rho(e_1)\ppair{e_4\circ{e_2},e_3}\\
\nonumber&&-\rho(e_2)\ppair{e_4\circ{e_3},e_1}+\rho(e_3)\ppair{e_4\circ{e_2},e_1}-\ppair{e_2\circ{e_3},e_1\circ{e_4}}\\
\nonumber&&+\ppair{e_1\circ{e_3},e_2\circ{e_4}}-\ppair{e_2\circ{e_3},e_4\circ{e_1}}-\ppair{e_1\circ{e_2},e_3\circ{e_4}}\\
&&+\ppair{e_4\circ{e_3},e_2\circ{e_1}}-\ppair{e_4\circ e_2,e_3\circ
e_1}.
\end{eqnarray*}
By   (iii) in Definition \ref{defi:preCA}, \eqref{eqn:skew} and the definition of $\huaD$, we
have
\begin{eqnarray*}
&&\rho(e_1)\ppair{e_2\circ{e_3},e_4}-\rho(e_1)\ppair{e_4\circ{e_2},e_3}\\
&=&\rho(e_1)\rho(e_2)\ppair{e_3,e_4}-\rho(e_1)\ppair{e_3,e_2\circ e_4}-\rho(e_1)\ppair{e_3,e_4\circ e_2}\\
&=&\rho(e_1)\rho(e_2)\ppair{e_3,e_4}-\rho(e_1)\rho(e_3)\ppair{e_2,e_4}.
\end{eqnarray*}
Similarly, we get
\begin{eqnarray*}
&&-\rho(e_2)\ppair{e_1\circ{e_3},e_4}-\rho(e_2)\ppair{e_4\circ{e_3},e_1}\\
&=&-\rho(e_2)\rho(e_1)\ppair{e_3,e_4}-\rho(e_2)\rho(e_4)\ppair{e_3,e_1}+\rho(e_2)\rho(e_3)\ppair{e_1,e_4},\\
&&\rho(e_3)\ppair{e_1\circ{e_2},e_4}+\rho(e_3)\ppair{e_4\circ{e_2},e_1}\\
&=&\rho(e_3)\rho(e_1)\ppair{e_2,e_4}
+\rho(e_3)\rho(e_4)\ppair{e_1,e_2}-\rho(e_3)\rho(e_2)\ppair{e_1,e_4},\\
&&-\rho(e_4)\ppair{e_1\circ{e_2},e_3}+\rho(e_4)\ppair{e_2\circ{e_3},e_1}\\&=&\rho(e_4)\rho(e_2)\ppair{e_1,e_3}
-\rho(e_4)\rho(e_3)\ppair{e_1,e_2},
\end{eqnarray*}
and
\begin{eqnarray*}
-\ppair{e_2\circ{e_3},e_1\circ{e_4}}-\ppair{e_2\circ{e_3},e_4\circ{e_1}}&=&-\rho(e_2\circ e_3)\ppair{e_1,e_4},\\
-\ppair{e_1\circ{e_2},e_3\circ{e_4}}+\ppair{e_4\circ{e_3},e_2\circ{e_1}}&=&\rho(e_4\circ
e_3)\ppair{e_1,e_2}
-\rho(e_1\circ e_2)\ppair{e_3,e_4},\\
\ppair{e_1\circ{e_3},e_2\circ{e_4}}-\ppair{e_4\circ{e_2},e_3\circ{e_1}}&=&\rho(e_2\circ
e_4)\ppair{e_1,e_3} -\rho(e_3\circ e_1)\ppair{e_2,e_4}.
\end{eqnarray*}
By  (i) in Definition \ref{defi:preCA}, it is straightforward to see $L=0$. Thus,  $J\in
\huaC^3 (E,\ker(\rho))$, and $J^\flat$ is well-defined.

Now we prove $\partial J=0$. By straightforward computations, we
have
\begin{eqnarray*}
\partial J&=&e_1\circ{J(e_2,e_3,e_4)}-e_2\circ{J(e_1,e_3,e_4)}+e_3\circ{J(e_1,e_2,e_4)}+{J(e_1,e_2,e_3)}\circ e_4\\
&&-J(e_1\circ{e_2},e_3,e_4)+J(e_1\circ{e_3},e_2,e_4)-J(e_1\circ{e_4},e_2,e_3)-J(e_2\circ{e_3},e_1,e_4)\\
&&+J(e_2\circ{e_4},e_1,e_3)-J(e_3\circ{e_4},e_1,e_2)\\
&=&e_1\circ(e_2\circ(e_3\circ{e_4}))-e_1\circ((e_2\circ
e_3)\circ{e_4})-e_1\circ(e_3\circ(e_2\circ{e_4}))\\
&&-e_2\circ(e_1\circ(e_3\circ{e_4}))+e_2\circ((e_1\circ
e_3)\circ{e_4})+e_2\circ(e_3\circ(e_1\circ{e_4}))\\
&&+e_3\circ(e_1\circ(e_2\circ{e_4}))-e_3\circ((e_1\circ
e_2)\circ{e_4})-e_3\circ(e_2\circ(e_1\circ{e_4}))\\
&&+J(e_1,e_2,e_3)\circ e_4\\
&&-(e_1\circ e_2)\circ(e_3\circ e_4)+((e_1\circ e_2)\circ e_3)\circ
e_4+e_3\circ((e_1\circ e_2)\circ
e_4)\\
&&+(e_1\circ e_3)\circ(e_2\circ e_4)-((e_1\circ e_3)\circ e_2)\circ
e_4-e_2\circ((e_1\circ e_3)\circ
e_4)\\
&&-(e_1\circ e_4)\circ(e_2\circ e_3)+((e_1\circ e_4)\circ e_2)\circ
e_3+e_2\circ((e_1\circ e_4)\circ
e_3)\\
&&-(e_2\circ e_3)\circ(e_1\circ e_4)+((e_2\circ e_3)\circ e_1)\circ
e_4+e_1\circ((e_2\circ e_3)\circ
e_4)\\
&&+(e_2\circ e_4)\circ(e_1\circ e_3)-((e_2\circ e_4)\circ e_1)\circ
e_3-e_1\circ((e_2\circ e_4)\circ
e_3)\\
&&-(e_3\circ e_4)\circ(e_1\circ e_2)+((e_3\circ e_4)\circ e_1)\circ
e_2+e_1\circ((e_3\circ e_4)\circ{e_2}).
\end{eqnarray*}
There are twenty-eight terms in all. Obviously, we have
\begin{eqnarray}
\nonumber&&-e_1\circ((e_2\circ e_3)\circ{e_4})+e_2\circ((e_1\circ
e_3)\circ{e_4})-e_3\circ((e_1\circ e_2)\circ{e_4})\\
    &&+e_3\circ((e_1\circ e_2)\circ
e_4) -e_2\circ((e_1\circ e_3)\circ e_4)+e_1\circ((e_2\circ e_3)\circ
e_4)=0.
\end{eqnarray}
On the other hand, by \eqref{eqn:skew} and \eqref{pro3}, we have
\begin{eqnarray*}
&&J(e_1,e_2,e_3)\circ e_4+((e_1\circ e_2)\circ e_3)\circ
e_4-((e_1\circ e_3)\circ e_2)\circ e_4+((e_2\circ e_3)\circ
e_1)\circ
e_4\\
&=&J(e_1,e_2,e_3)\circ e_4-J(e_1,e_2,e_3)\circ e_4\\
&=&0.
\end{eqnarray*}
Now we have cancelled ten terms. By \eqref{eqn:skew}, \eqref{pro3'}
and (iii) in Definition \ref{defi:preCA}, we have
\begin{eqnarray*}
e_1\circ{(e_2\circ(e_3,e_4))}+e_1\circ((e_3\circ e_4)\circ
e_2)&=&e_1\circ\huaD \ppair{e_2,e_3\circ
e_4}\\
&=&\huaD\rho(e_1)\ppair{e_2,e_3\circ
e_4}\\
&=&\huaD\big(\ppair{e_1\circ e_2,e_3\circ e_4}+\ppair{
e_2,e_1\circ(e_3\circ e_4)}\big).
\end{eqnarray*}
Similarly, we have
\begin{eqnarray*}
-e_1\circ(e_3\circ(e_2\circ{e_4}))-e_1\circ((e_2\circ e_4)\circ
e_3)&=&-e_1\circ\huaD g(e_3,e_2\circ
e_4)\\
&=&-\huaD\big(\ppair{e_1\circ e_3,e_2\circ e_4}+\ppair{
e_3,e_1\circ(e_2\circ
e_4)}\big),\\
e_2\circ(e_3\circ(e_1\circ{e_4}))+e_2\circ((e_1\circ e_4)\circ
e_3)&=&e_2\circ\huaD \ppair{e_3,e_1\circ
e_4}\\
&=&\huaD\big(\ppair{e_2\circ e_3,e_1\circ e_4}+\ppair{
e_3,e_2\circ(e_1\circ
e_4)}\big),\\
-e_2\circ(e_1\circ(e_3\circ{e_4}))+((e_3\circ e_4)\circ e_1)\circ
e_2&=&-\huaD
\ppair{e_2,e_1\circ(e_3\circ e_4)},\\
-((e_2\circ e_4)\circ e_1)\circ e_3 + e_3\circ(e_1\circ(e_2\circ
e_4))&=&\huaD\ppair{e_3,e_1\circ(e_2\circ e_4)},\\
((e_1\circ e_4)\circ e_2)\circ e_3 - e_3\circ(e_2\circ(e_1\circ
e_4))&=&-\huaD\ppair{e_3,e_2\circ(e_1\circ e_4)},\\
 -(e_1\circ
e_2)\circ(e_3\circ e_4)-(e_3\circ e_4)\circ(e_1\circ e_2)&=&-\huaD
\ppair{e_1\circ e_2,
e_3\circ e_4},\\
(e_1\circ e_3)\circ(e_2\circ e_4)+(e_2\circ e_4)\circ(e_1\circ
e_3)&=&\huaD \ppair{e_1\circ e_3, e_2\circ e_4},\\
-(e_1\circ e_4) \circ(e_2\circ e_3)-(e_2\circ e_3)\circ(e_1\circ
e_1)&=&-\huaD \ppair{e_1\circ e_4, e_2\circ e_3}.
\end{eqnarray*}
It is obvious that the sum of the eighteen terms in the left hand side  is
zero. Thus, we have $\partial J=0$ as well as $\huaD (J^\flat)=0$ by
Lemma \ref{lem:comm}. \qed

\section{Leibniz 2-algebras and Lie 2-algebras}

With previous preparation, in this section we show that, for a
pre-Courant algebroid $E$, there is an associated  Leibniz 2-algebra
naturally. If  using the skew-symmetric bracket,   a Lie 2-algebra
can be obtained.

Consider the complex
\begin{equation}\label{eqn:complex}
  \mathbb E:\Gamma(\ker(\rho))\stackrel{i}{\longrightarrow}\Gamma(E).
\end{equation}

Define degree-0 operation $l_2:\mathbb E\times\mathbb
E\longrightarrow\mathbb E$ by
\begin{equation}\label{eqn:defileib2}
  \left\{\begin{array}{rcll} l_2(e_1,e_2)&=& e_1\circ e_2 & \mbox{in~ degree-0},~\forall~e_1,e_2\in\Gamma(E)\\
  l_2(e,\kappa)&=& e\circ \kappa &
  \mbox{in~ degree-1},~\forall~e\in\Gamma(E),\kappa\in\Gamma(\ker(\rho))\\
  l_2(\kappa,e)&=& \kappa\circ e,&\mbox{in~ degree-1},~\forall~~e\in\Gamma(E),\kappa\in\Gamma(\ker(\rho))
   \end{array}\right.
\end{equation}
Define degree-1 operator $l_3:\wedge^3\mathbb E\longrightarrow
\mathbb E$ by
\begin{equation}\label{eqn:defileib3}
l_3(e_1,e_2,e_3)= J(e_1,e_2,e_3),
\quad\forall~e_1,e_2,e_3\in\Gamma(E)
\end{equation}

\begin{thm}\label{thm:Leibniz2}
  With the above notations, associated to any pre-Courant algebroid $E$, $(\mathbb E,l_2,l_3)$ is a
  Leibniz $2$-algebra, where $\mathbb E,l_2,l_3$ are given by
  \eqref{eqn:complex},
  \eqref{eqn:defileib2} and \eqref{eqn:defileib3}
  respectively.
\end{thm}
\pf By definition, it is straightforward to see that properties
$\rm(a1)-(a3), ~(b1)-(b4)$ are satisfied. By Theorem \ref{thm:J0},
property (c) is also satisfied. Thus, $(\mathbb E,l_2,l_3)$ is a
  Leibniz
 2-algebra. \qed

 \begin{rmk}
   The main contribution of this paper is showing that the Jacobiator in a pre-Courant algebroid satisfies some ``closed'' condition naturally, which leads to the study of the Pontryagin class and higher dimensional algebra structures. However, one can see that we do not use   (ii) in Definition \ref{defi:preCA} in the above construction. In fact, one can further obtain a weak Lie $2$-algebra \cite{Roytenbergweak} from a pre-Courant algebroid. We omit the study along this direction since it is not the main purpose of the paper.
 \end{rmk}



Let $(E,\ppair{\cdot,\cdot},\rho)$ be a Courant vector bundle with
two  pre-Courant algebroid structures  $\circ$  and
$\widetilde{\circ}$. It is easy to see that $\omega \in
\huaC^2 (E,\ker(\rho))$, where
\begin{equation}\label{xingbian}
 \omega(e_1, e_2) :=
e_1\widetilde{\circ}~ e_2- e_1\circ e_2,  \, ~~ \, ~~ \forall
{~e_1,e_2}\in{\Gamma(E)}.
\end{equation}
Conversely, we have

\begin{lem}
 Let $(E,\ppair{\cdot,\cdot},\rho,\circ)$  be a pre-Courant
 algebroid and  $\omega\in\Gamma(\Hom(\wedge^2E,E))$. Then
$(E,\ppair{\cdot,\cdot},\rho,\widetilde{\circ})$, where
$\widetilde{\circ}=\circ+\omega,$ is still a pre-Courant algebroid
if and only if $\omega\in \huaC^2 (E,\ker(\rho))$.
\end{lem}
\pf  We have $$\rho(e_1\widetilde{\circ} ~ e_2)=\rho(e_1\circ
e_2)+\rho(\omega(e_1,e_2))=[\rho(e_1),\rho(e_2)]+\rho(\omega(e_1,e_2)),$$
which implies that   (i) in Definition \ref{defi:preCA} is equivalent to that the image of
$\omega$ is contained in $\ker(\rho)$.   (ii) in Definition \ref{defi:preCA} is equivalent to
that $\omega$ is skew-symmetric.   (iii) in Definition \ref{defi:preCA} is equivalent to that
$\omega$ is totally skew-symmetric, i.e.
$\ppair{\omega(e_1,e_2),e_3}+\ppair{e_2,\omega(e_1,e_3)}=0.$ Thus,
the deformed operation $\widetilde{\circ}$ is a pre-Courant
algebroid structure if and only if $\omega\in
\huaC^2 (E,\ker(\rho))$. \qed\vspace{3mm}

Two  pre-Courant algebroid structures  $\circ$ and
$\widetilde{\circ}$ are called equivalent if there exists a 2-form $
\beta\in\Omega^2(M)$ such that
 $$\widetilde{\circ}=\circ+ \rho^*(d\beta).$$
More precisely, $e_1 \widetilde{\circ} e_2=e_1 \circ
e_2+\rho^*(d\beta(\rho(e_1),\rho(e_2),\cdot))$. In this case,
$B:=\rho^*\beta \in C^2 (E)$ is called a {\bf $B$-field}. Given
a {$B$}-field, we define a $B$-field transformation
$e^B:E\longrightarrow E$ by
\begin{eqnarray*}
e^B(e)=e+B^\sharp(e), \quad\forall e\in\Gamma(E).
\end{eqnarray*}
We see that equivalent structures are different by a $B$-field
transformation.
\begin{pro}
With the notations above, if~$~\widetilde{\circ}=\circ+
\rho^*(d\beta)$,  the bundle isomorphism $e^{B}: E \rightarrow
E$ satisfies
\begin{itemize}
\item[$\bullet$]  $e_1\widetilde{\circ}~ e_2 =  e^{-B}( e^{B}e_1\circ e^{B}e_2),  \, ~~ \, ~~ \forall
{~e_1,e_2}\in{\Gamma(E)}$;

\item[$\bullet$] $\ppair{e^{B}e_1,e^{B}e_2} = \ppair{e_1, e_2}$ \, ~ ~ and ~~~ \,   $\rho\circ e^{B} = \rho
$;

\item[$\bullet$] $\widetilde{J} =J$, i.e. the Jacobiator is
invariant under a B-field transformation.
   \end{itemize}
\end{pro}
It is easy to see that if $d\beta = 0$, then
$\widetilde{\circ}=\circ$ and $e^{B}$ is an automorphism of the
pre-Courant algebroid $(E,\ppair{\cdot,\cdot},\rho,\circ)$. To
classify more general deformations, we have
\begin{defi}Let $(E,\ppair{\cdot,\cdot},\rho)$ be a Courant vector bundle. Let $H_1$ and $H_2$ be
  sections of $C^{k+1}(E)$. We say that $H_1$ and $H_2$
 are equivalent if there exists a k-form $h\in\Omega^k(M)$ such that
 $$
 H_1-H_2=\rho^*(dh).
 $$
In particular,  the equivalence class $[J^\flat]$ of the Jacobiator
 $J^\flat  \in C^4(E)$ of a pre-Courant algebroid  is called the
{\bf Pontryagin class} of this pre-Courant algebroid.
\end{defi}

\begin{pro}
Let $(E,\ppair{\cdot,\cdot},\rho)$ be a Courant vector bundle with
two  pre-Courant algebroid structures  $\circ$, ~$\widetilde{\circ}$
and the corresponding Jacobiators $J$,  $\widetilde{J}$ respectively. Then one
has
$$ \circ - \widetilde{\circ}=\rho^*(h) ~~  \Longrightarrow~~~  J^\flat - \widetilde{J}^\flat =
\rho^*(dh), \mbox ~~ h\in\Omega^3(M).$$ That is, the two pre-Courant
algebroids have the same Pontryagin class if they can be deformed to
each other by a $3$-form on the base manifold.
\end{pro}
In particular, if the Pontryagin class of a pre-Courant algebroid
$(E,\ppair{\cdot,\cdot},\rho,\circ)$ vanishes, i.e.
$J^\flat=\rho^*(dh)$  then it is straightforward to see that  it can
be deformed into  a Courant algebroid with the new operation:
$\widetilde{\circ}=  \circ  - \rho^*(h)$. That is, the Pontryagin
class for a pre-Courant algebroid  is the {\em obstruction} to be
deformed into a Courant algebroid.

It is surprised that the Leibniz 2-algebra associated to an
arbitrary deformed pre-Courant algebroid is isomorphic to the
Leibniz 2-algebra associated to the original one. The follows is the
main result of this section.
\begin{thm}\label{thm:Leibniz2iso}
Let $(E,\ppair{\cdot,\cdot},\rho)$ be a Courant vector bundle.
$\circ$  and $\widetilde{\circ}$ are two  pre-Courant algebroid
structures with their associated Leibniz $2$-algebras  $\mathbb
  E$ and $\widetilde{\mathbb E}$ respectively. Then  $\mathbb
  E$ and $\widetilde{\mathbb E}$ are isomorphic.
\end{thm}

\pf Denote by $\widetilde{J}$ the Jacobiator of $\widetilde{\circ}=
\circ+\omega$, where $\omega\in \huaC^2 (E,\ker(\rho))$. Then by
straightforward computations, we have
\begin{equation}\label{eqn:equivalence}
 \widetilde{J}=J+\partial\omega+\half\omega^2,\end{equation} where
 $\omega^2:\Gamma(E)\times\Gamma(E)\times\Gamma(E)\longrightarrow\Gamma(E)$ is given by
 $$
\omega^2(e_1,e_2,e_3)=2\big(\omega(e_1,\omega(e_2,e_3))+c.p.\big).
 $$

Construct the isomorphism $(f_0,f_1,f_2)$ from $\mathbb E$ to
$\widetilde{\mathbb E}$ as follows: $f_0=\id:\Gamma(E)\longrightarrow
\Gamma(E)$, $f_1=\id:\Gamma(\ker(\rho))\longrightarrow
\Gamma(\ker(\rho))$ and
$f_2=\omega:\Gamma(E)\times\Gamma(E)\longrightarrow
\Gamma(\ker(\rho))$. It is obvious that \eqref{eqn:DGLA morphism c
1} is satisfied. By straightforward computations, the left hand side
of \eqref{eqn:DGLA morphism c 2} is equal to
$J+\partial\omega+\half\omega^2-\widetilde{J}$. By
\eqref{eqn:equivalence}, it is zero. Thus,
$(f_0={\id},f_1={\id},f_2=\omega):\mathbb E\longrightarrow
\widetilde{\mathbb E}$ is an isomorphism. \qed\vspace{3mm}

\emptycomment{

For a pre-Courant algebroid $E$, we consider the following deformed
 bracket operation:
\[e_1\circ_\omega e_2=e_1\circ e_2+\omega(e_1,e_2),\]
where $\omega\in\Gamma(\wedge^2E,E)$. It is easy to obtain:
\begin{pro}
With the above notations,
$(E,\ppair{\cdot,\cdot},\rho,\circ_\omega)$ gives a pre-Courant
algebroid if and only if $\rho\circ\omega=0$, and for any
$e_1,e_2,e_3\in\Gamma(E)$, the map $(e_1,e_2,e_3)\longmapsto
\ppair{\omega(e_1,e_2),e_3}$ is totally skew-symmetric with respect
to the pseudo-metric $\ppair{\cdot,\cdot}$.
\end{pro}
Suppose $\omega\in\Gamma(\wedge^2E,E)$ satisfies the conditions
above, it is obvious that $\omega\in C^2_\huaD(E,\Ker(\rho))$. Let
 $J'$ be the Jacobiator of the new pre-Courant algebroid, we have
\[J'=J+\partial\omega+J_\omega,\]
where $J_\omega$ is the Jacobiator of $\omega$, i.e.
$J_\omega(e_1,e_2,e_3)=\omega(e_1,\omega(e_2,e_3))+c.p..$

\begin{pro}
Let $\mathbb E$ and $\mathbb E_\omega $ be the corresponding Leibniz
2-algebra of the pre-Courant algebroids
$(E,\ppair{\cdot,\cdot},\rho,\circ)$ and
$(E,\ppair{\cdot,\cdot},\rho,\circ_\omega)$. If $J_\omega=0$,
Leibniz 2-algebras  $\mathbb E_\omega $ and $\mathbb E$ are
isomorphic.
\end{pro}

\pf Define $(f_0,f_1,f_2):\mathbb E_\omega\longrightarrow\mathbb E$
 by $f_0=\id,f_1=\id$ and $f_2=-\omega$. It is obvious that \eqref{eqn:DGLA morphism c
 1} holds. Since $J_\omega=0$, we have $J'=J+\partial\omega$, which
 is exactly \eqref{eqn:DGLA morphism c 2}. Thus,
 $(f_0=\id,f_1=\id,f_2=-\omega)$ is an isomorphism between Leibniz
 2-algebras. \qed\vspace{3mm}}

In \cite{Leibniz2}, it is shown  that if a twisted
Courant algebroid\footnote{We will see in next section that an exact
twisted Courant algebroid by a closed 4-form is exactly an exact
pre-Courant algebroid.}$E$ by a closed 4-form $H$ is exact, i.e. we have the following exact sequence
$$
0\longrightarrow T^*M \longrightarrow E\stackrel{\rho}{\longrightarrow } TM\longrightarrow 0,
$$
then $H$ must be exact, i.e.
$H=dh$, for some $h\in\Omega^3(M)$. The bracket operation is given
by
\begin{equation}
 \Dorfman{X+\xi,Y+\eta}=[X,Y]+L_X\eta-i_Yd\xi+h(X,Y).
\end{equation}
Thus, it can be seen as a deformation of the standard Courant
algebroid by  and element $\omega\in \huaC^2 (E,\ker(\rho))$, where $E=TM\oplus
T^*M$, $\rho$ is the projection, and $\omega$ is given by
$$
\omega(X+\xi,Y+\eta)=h(X,Y).
$$
\begin{cor}
  The Leibniz $2$-algebra associated to an exact pre-Courant
  algebroid is isomorphic to the strict Leibniz
  $2$-algebra associated to
  the standard Courant algebroid.
\end{cor}

Next we construct a Lie 2-algebra  associated to a pre-Courant
algebroid. For an $H$-twisted Courant algebroid,  such a Lie
2-algebra structure has been pointed out in \cite{GrutzmannC}. Since
we find that this is a nontrivial  construction, we shall give
detailed procedure below. Define the skew-symmetric bracket:
\begin{equation}\label{eq:skew-symmetrization}
  \Courant{e_1,e_2}=\half(e_1\circ e_2-e_2\circ e_1)=e_1\circ
  e_2-\half\huaD\ppair{e_1,e_2}.
\end{equation}
Denote by $\huaJ:\wedge^3\Gamma(E)\longrightarrow \Gamma(E)$ its
Jacobiator:
$$
\huaJ(e_1,e_2,e_3)= \Courant{e_1,\Courant{e_2,e_3}}+c.p..
$$
By means of  \cite[Proposition 2.6.5]{Roytenbergthesis}, we have
$$
\huaJ(e_1,e_2,e_3)=J(e_1,e_2,e_3)-\huaD T(e_1,e_2,e_3),
$$
where $T(e_1,e_2,e_3)$ is given by
$$
T(e_1,e_2,e_3)=\frac{1}{6}\big(\ppair{\Courant{e_1,e_2},e_3}+c.p.\big).
$$

Now we define a 2-term complex as follows:
\begin{equation}\label{eqn:defihuaE}
  \huaE:\Gamma(\ker(\rho))\stackrel{i}{\longrightarrow}\Gamma(E).
\end{equation}
Define degree-0 operation $l_2:\wedge^2\huaE\longrightarrow\huaE$ by
\begin{equation}\label{eqn:defil2}
  \left\{\begin{array}{rcll} l_2(e_1,e_2)&=& \Courant{e_1,e_2} & \mbox{in~ degree-0},~\forall~e_1,e_2\in\Gamma(E),\\
  l_2(e_1,\kappa)&=& \Courant{e_1,\kappa} &
  \mbox{in~
  degree-1},~\forall~e_1\in\Gamma(E),\kappa\in\Gamma(\ker(\rho)).
   \end{array}\right.
\end{equation}
Define degree-1 operator $l_3:\wedge^3\huaE\longrightarrow \huaE$ by
\begin{equation}\label{eqn:defil3}
 \begin{array}{rcll} l_3(e_1,e_2,e_3)&=& \huaJ(e_1,e_2,e_3), & \mbox{in~ degree-0},~\forall~e_1,e_2,e_3\in\Gamma(E).\\
   \end{array}
\end{equation}

\begin{thm}\label{thm:Lie3}
  For a pre-Courant algebroid $E$, $(\huaE,l_2,l_3)$ is a Lie
 $2$-algebra, where $\huaE,l_2,l_3$ are given by
  \eqref{eqn:defihuaE},
  \eqref{eqn:defil2}, \eqref{eqn:defil3}
  respectively.
\end{thm}
\pf It is obvious that we only need to show
\begin{eqnarray}
  \nonumber&&\sum_{i=1}^4(-1)^{i+1}\Courant{e_i,\huaJ(e_1,\cdots,\hat{e_i},\cdots,e_4)}\\
  \label{eqn:DJB}&&+\sum_{i<j}(-1)^{i+j}\huaJ(\Courant{e_i,e_j},e_1,\cdots,\hat{e_i},\cdots,\hat{e_j},\cdots,e_4)=0.
\end{eqnarray}
Write the left hand side of \eqref{eqn:DJB} as $K_1-K_2$, where
\begin{eqnarray*}
  K_1&=&\sum_{i=1}^4(-1)^{i+1}\Courant{e_i,J(e_1,\cdots,\hat{e_i},\cdots,e_4)}\\
&&+\sum_{i<j}(-1)^{i+j}J(\Courant{e_i,e_j},e_1,\cdots,\hat{e_i},\cdots,\hat{e_j},\cdots,e_4),
\end{eqnarray*}
and
\begin{eqnarray*}
K_2&=&\sum_{i=1}^4(-1)^{i+1}\Courant{e_i,\huaD T(e_1,\cdots,\hat{e_i},\cdots,e_4)}\\
  &&+\sum_{i<j}(-1)^{i+j}\huaD
  T(\Courant{e_i,e_j},e_1,\cdots,\hat{e_i},\cdots,\hat{e_j},\cdots,e_4).
\end{eqnarray*}

By straightforward computations, we have
\begin{eqnarray*}
K_1
&=&\nonumber\sum_{i=1}^4(-1)^{i+1}\big(e_i\circ J(e_1,\cdots,\hat{e_i},\cdots,e_4)-\half\ppair{e_i,J(e_1,\cdots,\hat{e_i},\cdots,e_4)}\big)\\
&&+\sum_{i<j}(-1)^{i+j}J(e_i\circ
e_j-\half\huaD\ppair{e_i,e_j},e_1,\cdots,\hat{e_i},\cdots,\hat{e_j},\cdots,e_4)\\
&=&\nonumber\sum_{i=1}^3(-1)^{i+1}\big(e_i\circ J(e_1,\cdots,\hat{e_i},\cdots,e_4)-\half\ppair{e_i,J(e_1,\cdots,\hat{e_i},\cdots,e_4)}\big)\\
&&+J(e_1,e_2,e_3)\circ e_4-\half\huaD\ppair{e_4,J(e_1,e_2,e_3)}\\
&&+\sum_{i<j}(-1)^{i+j}J(e_i\circ
e_j,e_1,\cdots,\hat{e_i},\cdots,\hat{e_j},\cdots,e_4)\\
&=&(\partial
J)(e_1,e_2,e_3,e_4)+\huaD\big(J^\flat(e_1,e_2,e_3,e_4)\big).
\end{eqnarray*}
By Theorem \ref{thm:J0}, we have $\partial J=0$, so we have
$K_1=\huaD\big(J^\flat(e_1,e_2,e_3,e_4)\big)$.

 In the case of Courant algebroids, by \cite[Theorem 2.4.3]{Roytenbergthesis}, we
 have $K_2=0$. Now for pre-Courant algebroids, after similar
 computations as in \cite[Theorem 2.4.3]{Roytenbergthesis}, we have
 $K_2=\huaD\big(J^\flat(e_1,e_2,e_3,e_4)\big)$. Thus, we have $K_1-K_2=0$. The proof is completed.
 \qed

 \begin{rmk}
    Even though a pre-Courant algebroid gives rise to a Lie $2$-algebra, it is not a Lie $2$-algebroid. From the  supergeometric point of view, as explained in \cite{Costa}, an arbitrary degree $3$ function $\Theta$  gives rise to a pre-Courant algebroid.  But in general $\{\Theta,\cdot\}$ is not a degree $1$ homological vector field. Thus, in general a pre-Courant algebroid is not a Lie $2$-algebroid.
 \end{rmk}

For any two pre-Courant algebroid structures  $\circ$  and
$ \circ'$ on a Courant vector bundle $(E,\ppair{\cdot,\cdot},\rho)$   with $\circ'=\circ+\omega$ for some
  $\omega \in
\huaC^2 (E,\ker(\rho))$, denote by $\Courant{\cdot,\cdot}'$ and $\Courant{\cdot,\cdot}$ their skew-symmetrization respectively. By \eqref{eq:skew-symmetrization}, we get
$$
\Courant{\cdot,\cdot}'=\Courant{\cdot,\cdot}+\omega.
$$
Similar to the proof of Theorem \ref{thm:Leibniz2iso}, we can show that
$(f_0={\id},f_1={\id},f_2=\omega):\huaE\longrightarrow
\huaE'$ is also an isomorphism between Lie 2-algebras.
\begin{pro}
The  associated Lie $2$-algebras to all possible
   pre-Courant algebroid structures on a Courant vector bundle
  are isomorphic to each other.
\end{pro}

\emptycomment{\section{Deformations of pre-Courant algebroids}

 Now let $(E,\ppair{\cdot,\cdot},\rho)$ be a Courant
vector bundle. $J_2$ and $J_1$ are the Jacobiators associated to the
pre-Courant algebroid structure $\circ_2$ and $\circ_1$
respectively.

\begin{defi}
$J_1^\flat,~J_2^\flat\in C^4_\huaD(E)$  are said to be equivalent,
if
\begin{equation}
  J_2^\flat=J_1^\flat+\huaD_1(\omega^\flat)+J_{\omega_1}^\flat,
\end{equation}
for some  $\omega_1\in C^2_\huaD(E,\Ker(\rho))$, where $\huaD_i$ is
the operator given by \eqref{eqn:defihuaD} associated to $\circ_i$
respectively, and $J_{\omega_1}$ is the Jacobiator for $\omega_1$,
i.e.
$J_{\omega_1}(e_1,e_2,e_3)=\omega_1(e_1,\omega_1(e_2,e_3))+c.p..$

For a pre-Courant algebroid $(E,\ppair{\cdot,\cdot},\rho,\circ)$
with the Jacobiator $J$, the equivalence class of $J^\flat$ is
called the Pontryagin class.
\end{defi}

We will see that (Proposition \ref{pro:strong}) in some special
case, the Pontryagin class of a pre-Courant algebroid is the pull
back of a closed 4-form on the base manifold, which is the
Pontryagin class of the induced ample Lie algebroid. This is why we
use ``the Pontryagin class'' here. See
\cite{firstP,dissection,Leibniz2} for more details.

\begin{pro}
 Let .
$J_3$, $J_2$ and $J_1$ are the Jacobiators associated to the
pre-Courant algebroid structure $\circ_3$, $\circ_2$ and $\circ_1$
respectively. If $J_3^\flat$ is equivalent to $J_2^\flat$ and
$J_2^\flat$ is equivalent to $J_1^\flat$, then $J_3^\flat$ is
equivalent to $J_1^\flat$. Thus, the above equivalence relation is
well-defined.
\end{pro}
\pf There exist  $\omega_i (i=1,2)\in C^2_\huaD(E,\Ker(\rho))$ such
that
\begin{eqnarray*}
  J_2^\flat=J_1^\flat+\huaD_1(\omega_1^\flat)+J_{\omega_1}^\flat,\quad
  J_3^\flat=J_2^\flat+\huaD_2(\omega_2^\flat)+J_{\omega_2}^\flat.
\end{eqnarray*}
Equivalently, we have
$$
 J_2=J_1+\partial_1(\omega_1)+J_{\omega_1},\quad
  J_3=J_2+\partial_2(\omega_2)+J_{\omega_2}.
$$
By Lemma \ref{lem:deform},
$$
\circ_3=\circ_2+\omega_2,\quad \circ_2=\circ_1+\omega_1.
$$
Therefore, we can obtain
$$
\partial_2(\omega_2)=\partial_1(\omega_2)+d_{\omega_1}\omega_2,
$$
where $d_{\omega_1}\omega_2$ is given by
$$
d_{\omega_1}\omega_2(e_1,e_2,e_3)=\omega_1(e_1,\omega_2(e_2,e_3))+\omega_2(e_1,\omega_1(e_2,e_3))+c.p..
$$

Then we have
\begin{eqnarray*}
  J_3&=&J_1+\partial_2(\omega_2)+\partial_1(\omega_1)+J_{\omega_1}+J_{\omega_2}=J_1+\partial_1(\omega_1+\omega_2)+J_{\omega_1+\omega_2},
\end{eqnarray*}
which implies that $J_3$ and $J_1$ are  equivalent. \qed}

\section{Regular pre-Courant algebroids }
As mentioned in the last section, in general, $\huaD^2 \neq 0$ since
the Jacobiator is nontrivial. The following proposition shows that
one can really get some cochain  complex from a kind of  pre-Courant
algebroids.
\begin{pro}\label{pro:d20}
  Let $E$ be a  pre-Courant algebroid such that
$$J(e_1,e_2,e_3) \in \rho^*(T^*M) = (\Ker(\rho))^\perp, ~~~ ~~ \forall
{~e_1,e_2, e_3}\in{\Gamma(E)}.$$ Then we have $\huaD^2=0$
  $(resp. ~\partial^2=0)$, i.e. $\huaD$ $(resp.~\partial)$ is a coboundary operator. Thus, $(C^\bullet(E),\huaD)$
  $( resp.~(\huaC^\bullet (E,\ker(\rho)),\partial))$ is a well-defined  cochain
  complex.
\end{pro}

  Follow \cite{Lodayalg}, we call the above cohomology the {\bf naive
  cohomology} of a pre-Courant algebroid. It will be interesting to
  explain  a pre-Courant algebroid using the language of super
  manifolds, and explain the naive cohomology correspondingly.

$E$ is said to be a {\bf regular pre-Courant algebroid}, if
$F:=\rho(E)$ has constant rank, i.e.  $F$ is a subbundle of $TM$. A pre-Courant algebroid $E$ is called a {\bf transitive pre-Courant algebroid}
when $\rho(E)=TM$.
 The next
proposition provides some criterions for Proposition \ref{pro:d20}.
\begin{pro}\label{pro:strong}
Let  $E$ be a regular pre-Courant algebroid. Then the  following
statements are equivalent:
\begin{itemize}
\item[\rm(S1)]
$J(e_1,e_2,e_3)\in{\Gamma((\Ker(\rho))^\perp)}$, ~~ $\forall
 {e_1,e_2,e_3}\in{\Gamma(E)}$;

\item[\rm(S2)] $J(\kappa,\cdot,\cdot)=0, $ ~~~ $\forall \kappa\in\Gamma(\Ker(\rho))$;

\item[\rm(S3)] There exists a $4$-form $H\in\Omega^4(M)$ closed along F-fiber, such that
$J^\flat=\rho^*H$.
\end{itemize}
\end{pro}

\pf $(S1)\Rightarrow(S2)$: If
$J(e_1,e_2,e_3)\in{\Gamma(\rho^*(T^*M))}=\Gamma((\ker(\rho))^\perp),
$
 for all ${e_1,e_2,e_3}\in{\Gamma(E)}$, then
 $J^\flat(e_1,e_2,e_3,\kappa)=0,$
for any ${e_1,e_2,e_3}\in\Gamma(E)$ and
$\kappa\in\Gamma(\ker(\rho))$. Since $J^\flat$ is skew-symmetric, so
we get
$$J^\flat(\kappa,e_2,e_3,e_1)=\ppair{J(\kappa,e_2,e_3),e_1}=0.$$ Since $\ppair{\cdot,\cdot}
$ is nondegenerate, we get $J(\kappa,\cdot,\cdot)=0,$ for all
$\kappa\in\Gamma(\ker(\rho))$.

$(S2)\Rightarrow(S3)$: if $J(\kappa,\cdot,\cdot)=0, $ for all $
\kappa\in\Gamma(\ker(\rho))$, we have
$$J^\flat(\kappa,\cdot,\cdot,\cdot,)=\ppair{J(\kappa,\cdot,\cdot),\cdot}=0, \quad
\forall~ \kappa\in\Gamma(\Ker(\rho)).$$ So there is a $4$-form $H\in
\Omega^4(M)$
 such that $J^\flat=\rho^*(H)$. It is not hard to deduce that
$$
\huaD J^\flat=\rho^*(dH).
$$
By Theorem \ref{thm:J0}, $\huaD J^\flat=0$. Thus, we have
$\rho^*(dH)=0$, which implies that $H$ is fibrewise closed.

$(S3)\Rightarrow(S1)$: For all ${e_1,e_2,e_3}\in\Gamma(E)$ and
$\kappa\in\Gamma(\Ker(\rho))$, we have
$$
\ppair{J(e_1,e_2,e_3),\kappa}=J^\flat(e_1,e_2,e_3,\kappa)=H(\rho(e_1),\rho(e_2),\rho(e_3),\rho(\kappa))=0.
$$
Thus, we have  $J(e_1,e_2,e_3)\in\Gamma((\Ker(\rho))^\perp)$. \qed

\begin{rmk}
  It is obvious that a twisted Courant algebroid by a closed $4$-form
  is a pre-Courant algebroid satisfying $J(e_1,e_2,e_3)\in{\Gamma((\Ker(\rho))^\perp)}$. By ${\rm(S3)}$, a pre-Courant
  algebroid satisfying $J(e_1,e_2,e_3)\in{\Gamma((\Ker(\rho))^\perp)}$ is not necessarily a twisted Courant algebroid by a
  closed $4$-form. There are only very tiny difference. In the transitive case, i.e. $\rho(E)=TM$, they are same. Since twisted Courant algebroids by
  closed $4$-forms arise naturally from the study of
three dimensional sigma models with Wess-Zumino term \cite{4form},
it would be interesting to investigate the physical meaning of
 pre-Courant algebroids.
\end{rmk}

In the sequel, a regular pre-Courant algebroid is called an {\bf $H$-twisted Courant algebroid} if it satisfies the equivalent conditions in Proposition \ref{pro:strong}.

\begin{rmk}
  For an $H$-twisted Courant algebroid, it is straightforward to see that we can construct a Leibniz $2$-algebra (Lie $2$-algebra) on the complex $\Gamma(\rho^*(T^*M))\stackrel{i}{\longrightarrow} \Gamma(E)$, which has the same
homotopy type as the Lie $2$-algebra $C^\infty(M)\stackrel{\huaD}{\longrightarrow}\Gamma(E)$ induced by a Courant algebroid.
\end{rmk}

\begin{cor}
Let $E$  be an $H$-twisted Courant algebroid, then
 $E/(\Ker(\rho))^\perp$ is a Lie algebroid.
\end{cor}
\pf If $E$ is  pre-Courant algebroid satisfying
$J(e_1,e_2,e_3)\in{\Gamma((\Ker(\rho))^\perp)}$, the induced
skew-symmetric bracket operation on $E/(\Ker(\rho))^\perp$ satisfies
the Jacobi identity, which implies that $E/(\Ker(\rho))^\perp$ is a
Lie algebroid. \qed\vspace{3mm}

  In the following, we will choose a dissection of
$E$, and get further insight to the results above.
Similarly to the treatment about Courant algebroids in
\cite{dissection}, we choose a dissection of a transitive
pre-Courant algebroid $E$ firstly, i.e. an isomorphism of vector
bundles:
\[\Psi:TM\oplus\mathcal{G}\oplus{T^*M}\longrightarrow{E}\] such that
\[\ppair{\Psi(x+r+\xi),\Psi(y+s+\eta)}=\xi(y)+\eta(x)+(r,s)^{\mathcal{G}},\] for all
$x,y\in\Gamma(TM),\xi,\eta\in\Gamma(T^*M)$ and
$r,s\in\Gamma(\mathcal{G})$, where $\mathcal{G}:={\Ker(
\rho)}/(\Ker(\rho))^\perp$ is a vector bundle with the induced
bilinear form $(\cdot,\cdot)^\mathcal{G}$ from
$\ppair{\cdot,\cdot}$. Following the methods in \cite{dissection},
dissections of a pre-Courant algebroid $E$ always exist.

Now fix a dissection
$\Psi:TM\oplus\mathcal{G}\oplus{T^*M}\longrightarrow{E}$, then the
isomorphism $\Psi$ transports the pre-Courant algebroid structure of
$E$ to $TM\oplus\mathcal{G}\oplus{T^*M}$. Given this dissection, the
Dorfman bracket $\circ$ induces four canonical maps:

$(a) \
\nabla:\Gamma(TM)\otimes\Gamma(\mathcal{G})\longrightarrow{\Gamma(\mathcal{G})}:$
\[\nabla_xr=Pr_\mathcal{G}(x\circ{r}),\ \ \ \forall{x\in\Gamma(TM),r\in\Gamma(\mathcal{G})};\]

$(b) \
R:\Gamma(TM)\otimes\Gamma(TM)\longrightarrow\Gamma(\mathcal{G}):$
\[R(x,y)=Pr_\mathcal{G}(x\circ{y}), \ \ \ \forall x,y\in\Gamma(TM);\]

$(c) \
[\cdot,\cdot]^\mathcal{G}:\Gamma(\mathcal{G})\otimes\Gamma(\mathcal{G})\longrightarrow\Gamma(\mathcal{G})$
\[[r,s]^\mathcal{G}=Pr_\mathcal{G}(r\circ{s})\ \ \ \ \forall{r,s\in\Gamma(\mathcal{G})};\]

$(d) \
\Phi:\Gamma(TM)\otimes\Gamma(TM)\otimes\Gamma(TM)\longrightarrow{C^\infty(M)}:$
\[\Phi(x,y,z)=Pr_{T^*M}(x\circ{y})(z), \ \ \ \ \forall x,y,z\in\Gamma(TM). \] By the properties of
a pre-Courant algebroid, it is easy to see that $R$ and $\Phi$ is
skew-symmetric and $C^\infty(M)$-bilinear. Using
$\nabla,R,\Phi,[\cdot,\cdot]^\mathcal{G}$, we could give the Dorfman
operator on $TM\oplus\mathcal{G}\oplus T^*M$: for all
$x_i\in\Gamma(TM),r_i\in\Gamma(\mathcal{G}),\xi_i\in\Gamma(T^*M),i=1,2,3,4$,
\begin{eqnarray*}
x_1\circ{x_2}&=&[x_1,x_2]+R(x_1,x_2)+\Phi(x_1,x_2,\cdot),\\
r_1\circ{r_2}&=&[r_1,r_2]^\mathcal{G}+\ppair{r_2,\nabla r_1},\\
x_1\circ{r_1}&=&-r_1\circ
x_1=\nabla_{x_1}r_1-\ppair{r_1,R(x_1,\cdot)},\\
x_1\circ{\xi_1}&=&L_{x_1}\xi_2,\\ \xi_1\circ x_1&=&-i_{x_1}d\xi_1,\\
r_1\circ\xi_1&=&\xi_1\circ r_1=\xi_1\circ\xi_2=0.
\end{eqnarray*}

And we can compute the Jacobiator directly:
\begin{eqnarray*}
J(x_1,x_2,x_3)&=&-\frac{1}{2}(R\wedge{R})^\mathcal{G}(x_1,x_2,x_3,\cdot)+d\Phi(x_1,x_2,x_3,\cdot)\\
&&+
\{\nabla_{x_1}R(x_2,x_3)-R([x_1,x_2],x_3))+c.p. \},\\
J(x_1,x_2,r_1)&=&\nabla_{x_1}\nabla_{x_2}r_1-\nabla_{x_2}\nabla_{x_1}r_1-\nabla_{[x_1,x_2]}r_1-[R(x_1,x_2),r_1]^\mathcal{G}+U(x_1,x_2,r_1),\\
J(x_1,r_1,r_2)&=&\nabla_{x_1}[r_1,r_2]^\mathcal{G}-[\nabla_{x_1}r_1,r_2]^\mathcal{G}-[r_1,\nabla_{x_1}r_2]^\mathcal{G}+V(x_1,r_1,r_2),\\
J(r_1,r_2,r_3)&=&[[r_1,r_2]^\mathcal{G},r_3]^\mathcal{G}+[[r_3,r_1]^\mathcal{G},r_2]^\mathcal{G}+[[r_2,r_3]^\mathcal{G},r_1]^\mathcal{G}+W(r_1,r_2,r_3),\\
J(\xi_1,e_1,e_2)&=&0,
\end{eqnarray*}
where $(R\wedge R)^\mathcal{G}$ is the $4$-form on $M$ given by
\begin{equation}
(R\wedge
R)^{\mathcal{G}}(x_1,x_2,x_3,x_4)=\frac{1}{4}\sum_{\sigma\in
S_4}sgn(\sigma)(R(x_{\sigma(1)},x_{\sigma(2)}),R(x_{\sigma(3)},x_{\sigma(4)}))^\huaG,
\end{equation}
and
$U\in\Gamma(\mathcal{G}^*\otimes\wedge^3T^*M),V\in\Gamma(\wedge^2\mathcal{G}^*\otimes\wedge^2T^*M),W\in\Gamma(\wedge^3\mathcal{G}^*\otimes
T^*M)$ are defined by:
\begin{eqnarray*}
U(x_1,x_2,r_1)(x_3)&=&-(\nabla_{x_1}R(x_2,x_3)-R([x_1,x_2],x_3))+c.p.,r_1)^\mathcal{G},\\
V(x_1,r_1,r_2)(x_2)&=&(\nabla_{x_1}\nabla_{x_2}r_1-\nabla_{x_2}\nabla_{x_1}r_1-\nabla_{[x_1,x_2]}r_1-[R(x_1,x_2),r_1]^\mathcal{G},r_2)^\mathcal{G},\\
W(r_1,r_2,r_3)(x_1)&=&-(\nabla_{x_1}[r_1,r_2]^\mathcal{G}-[\nabla_{x_1}r_1,r_2]^\mathcal{G}-[r_1,\nabla_{x_1}r_2]^\mathcal{G},r_3)^\mathcal{G}.
\end{eqnarray*}
We see that the Pontryagin class of $TM\oplus \mathcal{G}\oplus
T^*M$ is given by
\begin{eqnarray*}
B(x_1,x_2,x_3,x_4)&=&-\frac{1}{2}(R\wedge{R})^\mathcal{G}(x_1,x_2,x_3,x_4)+d\Phi(x_1,x_2,x_3,x_4),\\
B(x_1,x_2,x_3,r_1)&=&(\nabla_{x_1}R(x_2,x_3)-R([x_1,x_2],x_3))+c.p.,r_1)^\mathcal{G},\\
B(x_1,x_2,r_1,r_2)&=&(\nabla_{x_1}\nabla_{x_2}r_1-\nabla_{x_2}\nabla_{x_1}r_1-\nabla_{[x_1,x_2]}r_1-[R(x_1,x_2),r_1]^\mathcal{G},r_2)^\mathcal{G},\\
B(x_1,r_1,r_2,r_3)&=&(\nabla_{x_1}[r_1,r_2]^\mathcal{G}-[\nabla_{x_1}r_1,r_2]^\mathcal{G}-[r_1,\nabla_{x_1}r_2]^\mathcal{G},r_3)^\mathcal{G},\\
B(r_1,r_2,r_3,r_4)&=&([[r_1,r_2]^\mathcal{G},r_3]^\mathcal{G}+[[r_3,r_1]^\mathcal{G},r_2]^\mathcal{G}+[[r_2,r_3]^\mathcal{G},r_1]^\mathcal{G},r_4)^\mathcal{G},\\
B(\xi,\cdot,\cdot,\cdot)&=&0.
\end{eqnarray*}
Now if $J(e_1,e_2,e_3)\in{\Gamma((\Ker(\rho))^\perp)}$, we know that
the operator $\circ$ induces a Lie algebroid structure on the bundle
$E/(\Ker(\rho))^\perp=TM\oplus \mathcal{G}$, it is equivalent to say
that the maps $[\cdot,\cdot]^\mathcal{G},\nabla,R,\Phi$ satisfied the
following equalities:
\begin{eqnarray*}
[[r_1,r_2]^\mathcal{G},r_3]^\mathcal{G}+[[r_3,r_1]^\mathcal{G},r_2]^\mathcal{G}+[[r_2,r_3]^\mathcal{G},r_1]^\mathcal{G}&=&0,\\
\nabla_{x_1}[r_1,r_2]^\mathcal{G}-[\nabla_{x_1}r_1,r_2]^\mathcal{G}-[r_1,\nabla_{x_1}r_2]^\mathcal{G}&=&0,\\
\nabla_{x_1}R(x_2,x_3)-R([x_1,x_2],x_3))+c.p.&=&0 ,\\
\nabla_{x_1}\nabla_{x_2}r_1-\nabla_{x_2}\nabla_{x_1}r_1-\nabla_{[x_1,x_2]}r_1-[R(x_1,x_2),r_1]^\mathcal{G}&=&0.
\end{eqnarray*}
So when $J(e_1,e_2,e_3)\in{\Gamma((\Ker(\rho))^\perp)}$, we get
\begin{eqnarray*}
J(e_1,e_2,e_3)&=&-\frac{1}{2}(R\wedge{R})^\mathcal{G}(\rho(e_1),\rho(e_2),\rho(e_3),\cdot)+d\Phi(\rho(e_1),\rho(e_2),\rho(e_3),\cdot).
\end{eqnarray*}
By the Jacobiator given above, we see that $B=\rho^*H$ for a
$4$-form $H$ on  $M$ defined by $H=\frac{1}{2}
(R\wedge{R})^\mathcal{G}-d\Phi$. This is exactly the statement of
Proposition \ref{pro:strong} in the transitive case.

For a transitive quadratic Lie algebroid $A$, we can construct a
twisted Courant algebroid structure on $A\oplus T^*M$ following the
procedure above. Especially, let $\frkg$ be a quadratic Lie algebra with the Lie group $G$ its integration, and $P$  a $G$-principal bundle. With a dissection $\Psi:TM\oplus\mathcal{G}\oplus{T^*M}\longrightarrow TP/G\oplus{T^*M}$,
we have an extension of the Atiyah algebroid $TP/G$ to a twisted
Courant algebroid, and the twisted $4$-form is given by the first
Pontryagin class of $TP/G$ corresponding to the dissection. For more
details for the {\bf first Pontryagin class} of a quadratic Lie
  algebroid, see \rm\cite{firstP,dissection}.

\section{Construction of pre-Courant algebroids}
 Summarize the results of Izu Vaisman
\cite{Vai05}, we have a general method to construct pre-Courant
algebroids:
\begin{pro}\label{pro:construct}
For any Courant vector bundle $(E,\ppair{\cdot,\cdot},\rho)$, a pair
$(\nabla,\beta)$ gives a pre-Courant algebroid structure on $E$,
where $\nabla:\frkX(M)\times\Gamma(E)\longrightarrow\Gamma(E)$ is a
metric connection on $E$, i.e. $$X\ppair{e_1,e_2}=\ppair{\nabla
_Xe_1,e_2}+\ppair{e_1,\nabla _Xe_2},$$
 and $\beta\in\Gamma(\wedge^2E^*\otimes{E})$
satisfies the following properties:
\begin{itemize}
\item[$\bullet$] For any $e_1,e_2,e_3\in\Gamma(E)$, the map $(e_1,e_2,e_3)\longmapsto
\ppair{\beta(e_1,e_2),e_3}$ is totally skew-symmetric with respect
to the pseudo-metric $\ppair{\cdot,\cdot}$,

\item[$\bullet$]
$\rho(\beta(e_1,e_2))=[\rho(e_1),\rho(e_2)]-\rho(\nabla_{\rho(e_1)}e_2-
\nabla_{\rho(e_2)}e_1).$
\end{itemize}
Using the pair $(\nabla,\beta)$, the  bracket operation in the
pre-Courant algebroid is given by:
\[e_1\circ{e_2}=\nabla_{\rho(e_1)}e_2-
\nabla_{\rho(e_2)}e_1+\rho^*(\ppair{\nabla
e_1,e_2})+\beta(e_1,e_2),\quad\forall{e_1,e_2}\in\Gamma(E),\] where
$\rho^*(\ppair{\nabla e_1,e_2})$ is decided by:
$$\ppair{\rho^*\ppair{\nabla
e_1,e_2},e}=\ppair{\nabla_{\rho(e)}e_1,e_2},\quad\forall{e}\in\Gamma(E).
$$
\end{pro}
\pf  Using the properties of the pair $(\nabla,\beta)$, we can
verify that the operator $\circ$ defined above satisfies (i), (ii)
and (iii) in the definition of a pre-Courant algebroid.
\qed

\begin{rmk}
In the transitive case, given a dissection of a pre-Courant
algebroid $E$, $E=TM\oplus\mathcal{G}\oplus T^*M$. We can choose a
torsion free connection $\nabla^1$ on $TM$, and a connection
$\nabla^{2}$ on vector bundle $\mathcal{G}$ by
$\nabla^{2}_{x}r=Pr_\mathcal{G}x\circ r$. Then we get a metric
connection $\nabla$ on $E=TM\oplus\mathcal{G}\oplus T^*M$ by
$\nabla_x(y+r+\xi)=\nabla^1_xy+\nabla^{2}_xr+\nabla^1_x\xi$. We call
this connection $\nabla$ a good connection associated to a chosen
dissection of $E$. According to Proposition \ref{pro:construct},
suppose that the pre-Courant structure on $E$ is given by
$(\nabla,\beta)$, then compare the operator given in Proposition
\ref{pro:construct} and above, we have the following relationship:
\[\beta(x_1+r_1+\xi_1,x_2+r_2+\xi_2)=R(x_1,x_2)+\Psi(x_1,x_2,-)+[r_1,r_2]^g,\] for all
$x_1,x_2\in\Gamma(TM),r_1,r_2\in\Gamma(\mathcal{G}),\xi_1,\xi_2\in\Gamma(T^*M).$
\end{rmk}

It is known that one can construct  Courant algebroids from
coisotropic actions \cite{DLB}. In the following, we will define
twisted actions, and construct  pre-Courant algebroids by
coisotropic twisted actions.  First, let we recall how to construct
 Courant algebroids from  coisotropic actions in the language
of Proposition \ref{pro:construct}.

Let $(\frkg,[\cdot,\cdot]_\frkg,(\cdot,\cdot)^\frkg)$ be a quadratic
Lie algebra and $\rho:\frkg\rightarrow\frkX(M)$  a morphism. If
$\ker(\rho)$ is coisotropic, then the trivial bundle
$M\times{\frkg}$ is a Courant vector bundle with the anchor $\rho$. It
is well known that $M\times{\frkg}$  is an action Lie algebroid and
we denote the induced Lie bracket on $\Gamma(M\times \frkg)$ by
$[\cdot,\cdot]_{M\times\frkg}:$
$$
[e_1,e_2]_{M\times\frkg}=L_{\rho(e_1)}e_2-L_{\rho(e_2)}e_1+[e_1,e_2]_\frkg,
$$
where $[e_1,e_2]_\frkg$ is the pointwise Lie bracket of two sections
$e_1,e_2\in\Gamma(M\times\frkg)=C^\infty(M,\frkg)$. Define the
metric connection $\nabla$ by
$$\nabla_Xe=L_Xe,\quad X\in\frkX(M),~e\in\Gamma(M\times{\frkg})=C^\infty(M,\frkg),$$ and
define $\beta$ by $$\beta(e_1,e_2)=[e_1,e_2]_{\frkg}.$$ It is easy
to see the $(\nabla,\beta)$ satisfies the properties listed  in
Proposition \ref{pro:construct}, so that
$(M\times{\frkg},(\cdot,\cdot)^\frkg,\rho,\circ)$ is a pre-Courant
algebroid, where the operation $\circ$ is given by
\[e_1\circ e_2=L_{\rho(e_1)}e_2-L_{\rho(e_2)}e_1+[e_1,e_2]_{\frkg}+\rho^*(\nabla e_1,e_2)^\frkg.\]
 It is easy to check that $J=0$,  which implies that
$(M\times{\frkg},(\cdot,\cdot)^\frkg,\rho,\circ)$ is a Courant
algebroid.

Now, to get pre-Courant algebroids, we introduce the concept of
twisted actions as follows:

\begin{defi}\label{defi:action}
Let $\frkg$ be a Lie algebra and $(\rho,k)$ a pair of bundle maps:
$$\rho:M\times\frkg\longrightarrow TM; ~~~~\mbox ~~
k:\wedge^2(M\times\frkg)\longrightarrow M\times \frkg
$$
satisfying $k(e,\cdot)=0,$ ~~for all $e\in\ker(\rho)$ and
\begin{eqnarray}
\label{b}\rho([e_1,e_2]_{M\times\frkg})&=&[\rho(e_1),\rho(e_2)]-\rho(k(e_1,e_2)),
\quad \forall e_1,e_2\in \Gamma(M\times\frkg).
\end{eqnarray}
Then we call $(\rho,k)$  a twisted action of  $\frkg$  on $M$.

\end{defi}
If  we consider the twisted bracket $[\cdot,\cdot]_{k}$ on
$\Gamma(M\times\frkg)$:
$$
[e_1,e_2]_{k}=[e_1,e_2]_{M\times\frkg}+k(e_1,e_2).
$$
Equality \eqref{b} is saying that $\rho$ is a morphism with respect
to the twisted bracket $[\cdot,\cdot]_{k}$. But, in general,
$[\cdot,\cdot]_{k}$ is not a Lie bracket.  It seems interesting to
build the  framework for this kind of action.

\begin{thm}\label{actiontem}
Let $(\frkg,[\cdot,\cdot]_\frkg,(\cdot,\cdot)^\frkg)$ be a quadratic
Lie algebra with a  twisted action on $M$  by $(\rho,k)$. Then, if ~
$\ker(\rho)$ is coisotropic, there is a pre-Courant algebroid
structure on the bundle $M\times \frkg$  with the operation $\circ$
given by:
\[e_1\circ e_2=[e_1,e_2]_{M\times\frkg}+k(e_1,e_2)-\ppair{e_2,k(e_1,\cdot)}+\ppair{e_1,k(e_2,\cdot)}
+\rho^*(\nabla e_1,e_2)^\frkg.\]
\end{thm}
\pf First $\ker(\rho)$ being coisotropic means that $(M\times
\frkg,(\cdot,\cdot)^\frkg,\rho)$ is a Courant bundle. In the
following, we construct a pair $(\nabla,\beta)$ satisfying
conditions  in Proposition \ref{pro:construct}. We choose the metric
connection $\nabla$ on $M\times \frkg$  as follows:
$$
\nabla_Xe=L_Xe,\quad \forall
~X\in{\frkX(M)},~e\in\Gamma(M\times{\frkg})=C^\infty(M,\frkg).
$$
 By \eqref{b}, we have
\[
\rho[e_1,e_2]_\frkg+\rho(L_{e_1}e_2-
L_{e_2}e_1)=[\rho(e_1),\rho(e_2)]-\rho k(e_1,e_2).
\]
Therefore, we have
$$
\rho([e_1,e_2]_\frkg+k(e_1,e_2))=[\rho(e_1),\rho(e_2)]-\rho(L_{e_1}e_2-
L_{e_2}e_1) .
$$
Thus, $\widetilde{\beta}(e_1,e_2)=[e_1,e_2]_\frkg+k(e_1,e_2)$
satisfies the second condition in Proposition \ref{pro:construct}.
But $\widetilde{\beta}$ is not totally skew-symmetric with respect
to $(\cdot,\cdot)^\frkg$. Let $\beta$ be the skew-symmetrization of
$\widetilde{\beta}$ with respect to $(\cdot,\cdot)^\frkg$, i.e.
$$
\beta(e_1,e_2)=[e_1,e_2]_\frkg+k(e_1,e_2)-\ppair{e_2,k(e_1,\cdot)}+\ppair{e_1,k(e_2,\cdot)},
$$
where $\ppair{e_2,k(e_1,\cdot)},\ppair{e_1,k(e_2,\cdot)}\in
\Gamma(M\times \frkg)$ are given by
$$
\ppair{\ppair{e_2,k(e_1,\cdot)},e_3}=\ppair{e_2,k(e_1,e_3)},\quad
\ppair{\ppair{e_1,k(e_2,\cdot)},e_3}=\ppair{e_1,k(e_2,e_3)}.
$$
By the assumption for $k$,  we have
$$
\ppair{\ppair{e_2,k(e_1,\cdot)},\kappa}=\ppair{e_2,k(e_1,\kappa)}=0,\quad\forall \kappa\in\ker(\rho),
$$
which implies that
$$
\ppair{e_2,k(e_1,\cdot)}\in(\ker(\rho))^\perp\subset \ker(\rho).
$$
Similarly, we have $\ppair{e_1,k(e_2,\cdot)}\in  \ker(\rho)$. Thus,
$\beta$ defined above  satisfies the second condition in Proposition
\ref{pro:construct}. Therefore, $(\nabla,\beta)$ constructed above
gives a pre-Courant algebroid structure on $M\times\frkg$, and for
$e_1,e_2\in\Gamma(M\times \frkg)$ the operation $\circ$ is given by
\[e_1\circ e_2=[e_1,e_2]_{M\times\frkg}+k(e_1,e_2)-\ppair{e_2,k(e_1,\cdot)}+\ppair{e_1,k(e_2,\cdot)}
+\rho^*(\nabla e_1,e_2)^\frkg.\] The proof is finished.
\qed\vspace{3mm}

A nontrivial   example of twisted action comes from Cartan geometry (in particular, Parabolic geometry). We recall the
definition of Cartan geometry briefly (See \cite{weylstructure}).
\begin{defi}
Let $G$ be a Lie group with Lie algebra $\frkg$ and closed subgroup
$P$, a Cartan geometry $(\pi:\mathcal{G}\longrightarrow M,\omega)$
of type $(G,P)$ is  a principal $P$-bundle
$\mathcal{G}\longrightarrow{M}$ equipped a $\frkg$-valued $1$-form
(Cartan connection) satisfying the following conditions:
\begin{itemize}
\item[\rm(a)] the map $\omega_p:T_p\mathcal{G}\longrightarrow{\frkg}$ is a
linear isomorphism for every $p\in{\mathcal{G}}$;

\item[\rm(b)] ${r_a}^*\omega=Ad(a^{-1})\circ\omega$, for all  ${a}\in{P}$, where $r_a$ is  the right action of
 $a$ on $\mathcal{G}$;
\item[\rm(c)] $\omega( \hat{A})=A,$ for all ${A}\in{\frkp},$ where $\frkp$ is the Lie algebra of $P$ and  $\hat{A}$ is
 the fundamental vector field.
\end{itemize}
\end{defi}

The curvature of a Cartan connection is defined by the $\frkg$-valued
$2$-form
\begin{eqnarray}
K:=d\omega+\frac{1}{2}[\omega,\omega] ~~~
\in\Gamma(\wedge^2T\mathcal{G}\otimes\frkg).
\end{eqnarray}
Thus, the bundle  map $\omega^{-1}: M\times \frkg\longrightarrow
T\mathcal{G}$ satisfies
\begin{eqnarray}\label{Cartan}
\omega^{-1}([e_1,e_2]_{M\times\frkg})=[\omega^{-1}(e_1),\omega^{-1}(e_2)]-\omega^{-1}(K(\omega(e_1),\omega(e_2))),\quad \forall e_1,e_2\in \Gamma(M\times\frkg).
\end{eqnarray}
Consider the quadratic Lie algebra $\frkg \ltimes\frkg^*$ with the
Lie bracket:
\[ [A+\xi,B+\eta]=[A,B]_\frkg+\ad^*_A\eta-\ad^*_B\xi, \quad \forall A,B\in \frkg,\quad \forall \ \xi,\eta\in \frkg^*. \]
and the  invariant inner product: $$(A+\xi, B+\eta)_+ = \langle
A,\eta \rangle + \langle B,\xi \rangle.$$

 Next, for the  Cartan geometry $(\pi:\mathcal{G}\longrightarrow M,\omega)$
of type $(G,P)$, we shall define a  coisotropic twisted action of
$(\frkg \ltimes\frkg^*,(\cdot,\cdot)_+)$
 on $M$  by $(\rho,k)$. Set up
\[  \rho: \mathcal{G}\times(\frkg\ltimes\frkg^*)\longrightarrow
T\mathcal{G}; \mbox ~~~             \rho(A+\xi)\longrightarrow
\omega^{-1}(A),\]
and
\[  k:
\wedge^2(\mathcal{G}\times(\frkg\ltimes\frkg^*))\longrightarrow
\mathcal{G}\times(\frkg\ltimes\frkg^*); ~~~ \mbox ~~
k(A+\xi,B+\eta)= K(\omega^{-1}(A),\omega^{-1}(B))\]

It is not difficult  to see that $\ker(\rho)$ is coisotropic and
\[\rho([e_1,e_2]_{\mathcal{G}\times(\frkg\ltimes\frkg^*)})=[\rho(e_1),\rho(e_2)]-\rho(k(e_1,e_2)),
\quad \forall e_1,e_2\in \Gamma(\mathcal{G}\times(\frkg\ltimes
\frkg^*)).\]
Thus, $(\rho,k)$ is a coisotropic twisted action. By
Theorem {\rm\ref{actiontem}}, we give  a pre-Courant algebroid
structure on $\mathcal{G}\times(\frkg\ltimes \frkg^*)$.

Finally, we see that, as vector bundles
$$\mathcal{G}\times(\frkg\ltimes \frkg^*) \cong \mathcal{G}\times(\frkg\oplus \frkg^*)\cong T\mathcal{G}\oplus
T^*\mathcal{G}.$$ So that such a  pre-Courant algebroid  is actually
an exact twisted Courant algebroid.


\begin{thebibliography}{999}
\bibitem{ammardefiLeibnizalgebra}
M. Ammar and N. Poncin, Coalgebraic Approach to the Loday Infinity
Category, Stem Differential for $2n$-ary Graded and Homotopy
Algebras, \emph{Ann. Inst. Four.} 60 no. 1 (2010),  355-387.


\bibitem{Costa}
P. Antunes and J. M. Nunes da Costa, Hypersymplectic structures on Courant algebroids, \emph{J. Geom. Mechanics}  7 (3) (2015), 255-280.

\bibitem{parabolic}
S. Armstrong and R. Lu, Courant Algebroids in Parabolic Geometry,
arXiv:1112.6425.

\bibitem{baez:2algebras}
 J. Baez and A. S. Crans, Higher-Dimensional Algebra VI: Lie
 2-Algebras, \emph{Theory and Appl.  Categ.} 12 (2004),
 492-528.

 \bibitem{baezrogers}
  J. Baez and C. Rogers, Categorified Symplectic Geometry and the String Lie
  2-Algebra,   \emph{Homology, Homotopy and Applications} 12 (2010), 221-236.
\bibitem{DLB}
D. L. Bland and E. Meinrenken, Courant algebroid and Poisson
geometry, \emph{Int. Math. Res. Not. }, 11(2009),
2106-2145.

\bibitem{firstP}
P. Bressler, The first Pontryagin class, \emph{Compos. Math.} 143
(2007), no 5, 1127-1163.

\bibitem{CA}
P. Bressler and A. Chervov. Courant algebroids. \emph{J. Math. Sci.} (N. Y.), 128 (4) (2005), 3030-3053.



\bibitem{weylstructure}
A. \v{C}ap and J. Slov$\acute{a}$k, \emph{Parabolic geometry. I Background
and General Theorey}, Mathematical Surveys and Monographs, vol. 154,
American Mathematical Society, Providence, RI, 2009.



\bibitem{CLSecourant}
Z. Chen, Z. Liu and Y. Sheng, $E$-Courant algebroids, \emph{Int.
Math. Res. Not.} 22(2010), 4334-4376.

\bibitem{dissection}
Z. Chen, M. Stienon and P. Xu, On regular Courant algebroids,
\emph{J. Symplectic Geom.} 11 (2013), no. 1, 1–24.

\bibitem{Grutzmann}
M. Gr{\"u}tzmann, $H$-twisted Lie algebroids, \emph{J. Geom. Phys.}
61 (2011), 476-484.
\bibitem{GrutzmannC}
M. Gr{\"u}tzmann, $H$-twisted Courant algebroids, arXiv:1101.0993.

\bibitem{GS}
 M. Gr{\"u}tzmann and T. Strobl,  General Yang-Mills type gauge theories for $p$-form gauge fields: from physics-based ideas to a mathematical framework or from Bianchi identities to twisted Courant algebroids, \emph{Int. J. Geom. Methods Mod. Phys.} 12 (2015), no. 1, 1550009, 80 pp.

\bibitem{GualtieriGeneralizedComplex}
M. Gualtieri, \emph{Generalized Complex Geometry,} PhD thesis, St
John's College, University of Oxford, Nov. 2003.


\bibitem{4form}
M. Hansen and T. Strobl, First Class Constrained Systems and
Twisting of Courant Algebroids by a Closed 4-form, \emph{In
Fundamental Interactions, A Memorial Volume for Wolfgang Kummer,}
pages 115--144. World Scientific, 2010.

\bibitem{Hull}
 C. M. Hull, Generalised Geometry for M-Theory, \emph{J. High Energy Phys.} 7 (2007), 079.

 \bibitem{Jot}
M. Jotz Lean, N-manifolds of degree $2$ and metric double vector bundles, arXiv:1504.00880.

\bibitem{Yvettehistory}
Y. Kosmann-Schwarzbach, Courant algebroids. A short history. \emph{SIGMA Symmetry Integrability Geom. Methods Appl.} 9 (2013), Paper 014, 8 pp.


\bibitem{stasheff2} T. Lada and J. Stasheff, Introduction to sh Lie algebras for
physicists, \emph{Int. Jour. Theor. Phys.} 32 (7) (1993), 1087-1103.

\bibitem{Libland}
D. Li-Bland, AV-Courant algebroids and generalized CR structures,
\emph{Canad. J. Math.} 63 (2011), no. 4, 938–960.

\bibitem{lwx}%
Z.-J. Liu, A. Weinstein and P. Xu, Manin triples for Lie
bialgebroids, \emph{J. Diff. Geom.} 45 (1997), 547-574.

\bibitem{livernet}
M. Livernet, Homologie des alg$\rm\grave{e}$bres stables de matrices sur une $A_\infty$-alg$\rm\grave{e}$bre, \emph{C. R. Acad. Sci. Paris S$\rm\acute{e}$r. I
Math.}, 329(2) (1999), 113-116.

\bibitem{Loday and Pirashvili}
J.-L. Loday and T. Pirashvili, Universal enveloping algebras of
Leibniz algebras and (co)homology, \emph{Math. Ann.} 296 (1993),
139-158.




\bibitem{Roytenbergthesis}
D. Roytenberg, \emph{Courant algebroids, derived brackets and even
symplectic supermanifolds}, PhD thesis, UC Berkeley, 1999,
arXiv:math.DG/9910078.

\bibitem{roytenbergshl}
D. Roytenberg and A. Weinstein, Courant algebroids and strongly homotopy Lie
algebras, \emph{Lett. Math. Phys.}, 46 (1) (1998), 81-93.

\bibitem{Roytenbergweak} D. Roytenberg,  On weak Lie $2$-algebras,  XXVI Workshop on Geometrical Methods in Physics. \emph{ AIP Conf. Proc.}, vol. 956. 180-198. Amer. Inst. Phys., Melville, NY, 2007.

\bibitem{Leibniz2} Y. Sheng, Z. Liu, Leibniz 2-algebras and twisted Courant algebroids,
\emph{Comm. Algebra.} 41 (05) (2013), 1929-1953.





\bibitem{Lodayalg} M. Stienon and P. Xu, Modular classes of Loday algebroids.
\emph{C. R. Math. Acad. Sci. Paris} 346 (2008), no. 3-4, 193-198.

\bibitem{UchinoshL}
K. Uchino, Derived brackets and sh Leibniz algebras, \emph{J. Pure
Appl. Algebra}, 215 (2011) 1102-1111.


\bibitem{Vai05} I. Vaisman, Transitive Courant algebroids. \emph{Int. J. Math.
Math. Sci.}, 11 (2005) 1737-1758.


\bibitem{xiaomeng}
X. Xu, Twisted Courant algebroids and coisotropic Cartan geometries. \emph{J. Geom. Phys.} 82 (2014), 124–131.




\bibitem{zambon}
M. Zambon, $L_\infty$-algebras and higher analogues of Dirac
structures and Courant algebroids, \emph{J. Symplectic Geom.} 10 (2012), no. 4, 563–599.
\end{thebibliography}
\end{document}